\documentclass[aps,prd,twocolumn,superscriptaddress,preprintnumbers,nofootinbib,longbibliography,floatfix]{revtex4-1}

\usepackage{amssymb}
\usepackage{graphicx}
\usepackage{colortbl}
\usepackage{hyperref}
\usepackage{longtable}
\usepackage{booktabs}
\usepackage{array}
\usepackage{tikz}
\usepackage{tabularx}
\usepackage{enumitem}
\usepackage{outlines}
\usepackage{nicefrac}
\usepackage{makecell}
\usepackage[normalem]{ulem}
\usepackage{color}

\newcommand{\BBN}{\text{BBN}}
\newcommand{\CNN}{\text{CNN}}

\newcommand{\HLN}{\text{HLN}}
\newcommand{\PTM}{\text{HLN}_0}

\newcommand{\DO}{\text{DO}}
\newcommand{\TO}{\text{TO}}
\newcommand{\NN}{\text{NN}}

\newcommand{\HL}{\text{HL}}

\newcommand{\ADO}{\text{ADO}}
\newcommand{\AUC}{\text{AUC}}
\newcommand{\EFP}{\text{EFP}}

\usepackage{amsmath}

\DeclareMathOperator*{\argmax}{argmax}

\newcommand{\rref}[1]{Ref.~\cite{#1}}
\newcommand{\rrefs}[1]{Refs.~\cite{#1}}
\newcommand{\Fig}[1]{Fig.~\ref{#1}}

\newcommand{\Tab}[1]{Table~\ref{#1}}

\newcommand{\Sec}[1]{Sec.~\ref{#1}}
\newcommand{\Secs}[2]{Secs.~\ref{#1} and \ref{#2}}
\newcommand{\App}[1]{App.~\ref{#1}}

\newcommand{\Eq}[1]{Eq.~\eqref{#1}}
\newcommand{\Eqs}[2]{Eqs.~\eqref{#1} and \eqref{#2}}

\newcommand{\tndk}[3]{
	\begin{gathered}\includegraphics[width=0.035\textwidth]{figures_ndk_efp_#1_#2_#3.pdf}\end{gathered}
}

\newcommand{\ndk}[4]{
	\begin{gathered}\includegraphics[width=#1\textwidth]{figures_ndk_efp_#2_#3_#4.pdf}\end{gathered}
}

\begin{document}
\preprint{MIT--CTP 5223}

\title{Mapping Machine-Learned Physics into a Human-Readable Space}

\begin{abstract}
	We present a technique for translating a black-box machine-learned classifier operating on a high-dimensional input space into a small set of human-interpretable observables that can be combined to make the same classification decisions. We iteratively select these observables from a large space of high-level discriminants by finding those with the highest decision similarity relative to the black box, quantified via a metric we introduce that evaluates the relative ordering of pairs of inputs. Successive iterations focus only on the subset of input pairs that are misordered by the current set of observables. This method enables simplification of the machine-learning strategy, interpretation of the results in terms of well-understood physical concepts, validation of the physical model, and the potential for new insights into the nature of the problem itself. As a demonstration, we apply our approach to the benchmark task of jet classification in collider physics, where a convolutional neural network acting on calorimeter jet images outperforms a set of six well-known jet substructure observables. Our method maps the convolutional neural network into a set of observables called energy flow polynomials, and it closes the performance gap by identifying a class of observables with an interesting physical interpretation that has been previously overlooked in the jet substructure literature.
\end{abstract}

\author{Taylor Faucett}
\affiliation{Department of Physics and Astronomy, UC Irvine, Irvine, CA 92627}
\author{Jesse Thaler}
\affiliation{Center for Theoretical Physics, Massachusetts Institute of Technology, Cambridge, MA 02139}
\affiliation{The NSF AI Institute for Artificial Intelligence and Fundamental Interactions}
\author{Daniel Whiteson}
\affiliation{Department of Physics and Astronomy, UC Irvine, Irvine, CA 92627}
\date{\today}
\maketitle

{
\small
\tableofcontents
}

\section{Introduction}
\label{sec:intro}

It is widely appreciated that neural networks (NNs) and related machine-learning (ML) tools can provide powerful solutions to important and difficult problems in high-energy physics~\cite{Guest:2018yhq,Radovic:2018dip}. Examples of tasks that have benefitted from NNs include triggering~\cite{Kohne:1997ph}, event reconstruction~\cite{Aad:2014yva,Peterson:1988gs}, object identification~\cite{Abramowicz:1995zi,Khachatryan:2014ira,Abazov:2004vd}, and event selection~\cite{ABREU1992383,Abazov2001282,Aaltonen:2009jg}. In all of these contexts, though, the physicist wonders:  \emph{what has the machine learned?} Satisfaction with improved performance is tempered by frustration with the ``black-box'' nature of NN strategies.

The advent of deep learning has made this question more urgent, as the data dimensionality of the tasks has increased dramatically and ML approaches have outperformed human-engineered strategies for problems that, until recently, were deemed too difficult for ML. Examples include event classification~\cite{Baldi:2014kfa,Baldi:2014pta,Santos:2016kno,Aurisano:2016jvx,Cohen:2017exh,Andrews:2018gew}, jet substructure studies~\cite{Almeida:2015jua,deOliveira:2015xxd,Baldi:2016fql,Larkoski:2017jix,Kasieczka:2017nvn}, jet flavor classification~\cite{Guest:2016iqz,ATL-PHYS-PUB-2017-003,CMS-DP-2017-005,Sirunyan:2017ezt,Komiske:2016rsd}, detector unfolding~\cite{Andreassen:2019cjw,Datta:2018mwd,Bellagente:2020piv}, and uncertainty estimation~\cite{Englert:2018cfo,Barnard:2016qma,Bollweg:2019skg,Nachman:2019dol,Kasieczka:2020vlh}. In each case, a deep NN has successfully utilized more information from the high-dimensional low-level input data than was captured by a smaller number of physics-motivated high-level (HL) observables. When there is a performance gap between machine-learned and human-engineered strategies, physicists want to understand how the NN is using the low-level information to improve its performance. This desire also applies to situations where performance of the ML solution matches (but does not exceed) the human-engineered approach. Has the machine learned the same strategy that humans devised, or has it found an alternative solution with equal efficacy?

In this paper, we present a technique for translating a black-box ML strategy based on low-level inputs into a human-readable space of HL observables. Instead of trying to directly interpret an NN classifier, our approach is to use the NN as a guide for the construction of a simpler classifier which makes the same decisions but relies on a small number of physics-inspired human-interpretable inputs, selected iteratively from a large space of observables. As a demonstration of our method, we present a case study in jet classification~\cite{Baldi:2016fql}, using a convolutional neural network (CNN) to guide the selection of a small set of HL observables called \emph{energy flow polynomials} (EFPs)~\cite{Komiske:2017aww}. We find that the final set of HL observables provides the same classification performance as the CNN acting on the low-level inputs, but in a more compact, interpretable, and physically meaningful format. Our study suggests that physicists should consider an overlooked set of HL observables that are relevant to jet substructure classification.

This desire to gain insight into the nature of ML strategies is important on several levels. First, it is critical that information used by the NNs be validated as real and physical, rather than an artifact of the training samples or procedure. Even in cases where the ML training does not rely on simulated samples~\cite{Metodiev:2017vrx,Andreassen:2018apy,Collins:2018epr}, it is important to understand what information is being used. Translating a  black-box ML strategy into a simpler network that relies on a smaller number of physically meaningful observables allows for effective validation. Second, having an interpretable strategy based on HL inputs allows for more reliable estimates of systematic uncertainties. If the HL space is small enough, the HL inputs themselves can be individually studied and calibrated, which is a more straightforward task than handling low-level inputs directly. Third, if one can replace a sophisticated ML strategy with a simpler network based on preprocessed inputs, this can enable faster interference and lower memory requirements at run-time~\cite{10.1145/1150402.1150464,Duarte:2018ite}. Finally, if previously overlooked information in the low-level data is physical, identifying it can provide new insights into the nature of the problem. Indeed, in our jet classification case study, we show how the six HL observables studied in \rref{Baldi:2016fql} can be augmented by a seventh HL observable that had not been previously considered in the literature. By scrutinizing the structure of the EFPs, we can provide a physical interpretation for this new observable.

There have been previous proposed strategies to draw connections between a learned NN strategy and existing HL observables. This can done by comparing the performance with and without the HL observables~\cite{Chang:2017kvc} or projecting the decision surfaces along those observables~\cite{Baldi:2014kfa}. An alternative strategy is to expand the NN function in a basis of the input features~\cite{2016arXiv161200410A,Wunsch:2018oxb,Roxlo:2018adx}. These strategies are valuable, but are primarily limited to studying the structure of the NN in terms of already-identified  HL observables. The goal of the present work is to identify new HL observables relevant for ML tasks, starting from a large space of HL observables that is as broad and comprehensive as possible (yet still interpretable) and systematically mapping a black-box ML strategy into that space.

Our mapping strategy suggests a new approach to the application of deep learning to high-energy physics data. In this approach, training a powerful deep neural network (DNN) on low-level inputs is just the first step, which helps gauge the effective upper limit on possible ML performance and determine asymptotically optimal decision boundaries. The new second step is translating as much of the ML strategy as possible to a well-understood set of HL observables. This allows for physical interpretation of the information being used, validation of the modeling, definition of reasonable systematic uncertainties, as well as computational benefits due to dimensionality reduction.

An extended outline of this paper is as follows. In \Sec{sec:strategy}, we present a general approach to map an ML model's learned solution into a human-readable space. This mapping requires that we construct a large space of candidate HL observables and use a reliable similarity metric for comparing these observables to the learned solution. As our similarity metric, we introduce \emph{average decision ordering} (ADO), which is related to Kendall's rank correlation coefficient~\cite{ktau} and quantifies to what extent two classifiers make the same (even if incorrect) decisions. We then present a mapping strategy that leverages the ADO. Our \emph{black-box guided strategy} iteratively finds the maximum ADO between the HL observables and a fixed black-box ML algorithm.

In \Sec{sec:casestudy}, we review the collider task of discriminating between jets originating from boosted $W$ bosons and those originating from light quarks and gluons. This is a well-studied problem in the field of jet substructure~\cite{Seymour:1991cb,Seymour:1993mx,Butterworth:2002tt,Butterworth:2007ke,Butterworth:2008iy,Abdesselam:2010pt,Altheimer:2012mn,Shelton:2013an,Altheimer:2013yza,Adams:2015hiv,Larkoski:2017jix,Asquith:2018igt,Marzani:2019hun}, where both HL~\cite{Cui:2010km,Thaler:2010tr,Thaler:2011gf,Larkoski:2013eya,Aad:2014haa,Larkoski:2014gra} and NN strategies~\cite{deOliveira:2015xxd,Almeida:2015jua,Baldi:2016fql} have proven effective. Our starting point will be the analysis of \rref{Baldi:2016fql}, which found a small but persistent improvement in classification performance with a deeply-connected CNN when compared to a boosted decision tree (BDT) of HL observables. To augment the set of HL observables for jet tagging, we search the space of EFPs~\cite{Komiske:2017aww}, which form an (over)complete basis of collider observables that are infrared and collinear (IRC) safe. We also introduce EFP variants inspired by IRC-unsafe observables that have been successful in other jet tagging studies~\cite{Pandolfi:2012ima,Chatrchyan:2012sn,Larkoski:2014pca,Gras:2017jty}.

The results of our case study are presented in \Secs{sec:supplementing}{sec:iteratively}. We start with the six HL observables from \rref{Baldi:2016fql}, and use the black-box guided strategy to identify a seventh HL observable that closes the performance gap with the CNN. We then apply the black-box guided strategy starting from just the mass and transverse momentum of the jet, comparing the results to a brute force strategy of directly searching the space of EFPs and a guided search based on ground truth labels. The black-box guided strategy significantly outperforms the label guided search, reaching comparable performance to the brute force strategy with considerably reduced computational costs. At the end of each of these sections, we provide a physical interpretation of the translated ML strategy, and we draw broader conclusions in \Sec{sec:discussion}.

\section{Translating from Machine to Human}
\label{sec:strategy}
In our mapping approach, we seek to identify a small set of physically-motivated HL observables that, when combined into a joint classifier, make the same classification decisions as a deep network operating on the low-level  features. Crucially, our set of  HL features is designed to, when combined, maximize the classification performance by following the learned strategy of the black-box NN, which we argue below is more efficient than training directly on ground truth information. If this translation is successful, then we will have expressed the ML strategy more simply and transparently in terms of a few HL observables.

The first step in our approach is to identify a comprehensive set of HL observables that are potentially relevant for solving the ML task at hand. This in turn requires (human) knowledge about the physical system being studied and about the kinds of HL observables that have interpretable meaning. We know of no way to automate this step, though automation may not be desirable, as the choice and structure of the HL observable space defines the physical interpretation. For our jet substructure classification case study, the EFPs have already been identified as a suitable basis of HL observables~\cite{Komiske:2017aww}, as discussed further in \Sec{subsec:casestudy_efps}.\footnote{Beyond jet substructure, the EFPs are also formally a complete basis for event-wide IRC-safe observables, but their utility in that context has not yet been established.} Other ML tasks in high-energy physics might require the development of alternative bases of HL observables.

In the rest of this section, we describe the aspects of our approach that are generic to any ML task, focusing on the case of binary classification. To evaluate the relative performance of our simple HL network and a black-box NN, we need a metric to evaluate whether two classifiers make the same classification decisions. There are many such metrics one could use, but we introduce ADO, in part because it shares the conceptual simplicity of the area under the curve (AUC) metric often used to benchmark classifiers against ground truth. Armed with an explorable set of HL observables and a metric for assessing learning similarity, we then present a guiding strategy to map black-box NNs into a physically meaningful space.

\subsection{Average Decision Ordering}
\label{subsec:strategy_ado}
Our guided strategy requires a similarity metric that compares the output of two decision functions $f(x)$ and $g(x)$. Here, $x$ represents the full high-dimensional low-level  data, which are inputs to both the black-box NN and the physically-motivated HL observables. Of course, the definition of similarity must reflect the task for which these decision functions are applied, which in this case is binary classification. Because classification performance is invariant under any non-linear monotonic transformation of $f$ or $g$, our similarity metric cannot be affected by such a transformation. This rules out naive metrics like functional overlap or linear correlation.

Furthermore, it is not sufficient to simply compare the overall performance of two classifiers over a given dataset, since that does not provide information about how the low-level  inputs are being used. As discussed in \rref{Larkoski:2014pca}, two decision functions might use information from different regions of the low-level  input space and make conflicting classification decisions case by case, yet still achieve similar overall performance. The key to our guided strategies is that we aim to match not just the classification performance of the black-box NN, but also the specific classification decisions.

We assume that the decision functions $f(x)$ and $g(x)$ are real valued and that the final binary classification is determined by a threshold on the decision function output. Objects on one side of the threshold are labeled ``signal'', while objects on the other side are labeled ``background''. Depending on the particular application, this threshold can be tuned to different points on the receiver operating characteristic (ROC) curve to optimize the signal acceptance versus background rejection. To quantify overall classification performance, we use the AUC, which is equivalent to the probability that a randomly selected signal/background pair is correctly ordered by a decision function $f(x)$:
\begin{equation}
	\label{eq:AUC}
	\AUC[f]  = \int \mathrm{d}x\, \mathrm{d}x'\, p_{\text{sig}}(x) \, p_{\text{bkg}} (x') \, \Theta \big( f(x) - f(x') \big).
\end{equation}
Here, $\Theta$ is the Heaviside theta function (i.e.~$\Theta(x<0) =  0$ and $\Theta(x\ge 0) = 1$) and $p_{\text{sig}}$ and $p_{\text{bkg}}$ are the ground truth signal and background probability distributions. A perfect decision function has $\AUC = 1$ and random guessing yields $\AUC = \frac{1}{2}$.

To compare the classification behavior of two decision functions, we consider the \emph{decision surface} defined by a threshold on the function output. For each function, the set of such thresholds defines a set of surfaces in $x$ space. If two decision functions have the same decision surfaces, then they are effectively using the same low-level  information for classification. Note that the absolute output values of the decision functions are not relevant for determining whether the decision surfaces are similar. The relative locations of the decision surfaces are determined by the relative ordering of the two decision functions when evaluated at pairs of points in the input space. We can encapsulate this information via the \emph{decision ordering} (DO) for a pair of points $x$ and $x'$:
\begin{equation}
\label{eq:DO_def}
	\DO[f,g](x,x') = \Theta \Big( \big( f(x) - f(x') \big) \big( g(x) - g(x') \big) \Big),
\end{equation}
where 1 corresponds to $f$ and $g$ having the same ordering and 0 corresponds to inverted ordering.

If two decision functions have $\DO = 1$ for all pairs of $x$ and $x'$, then they are monotonically related to each other, have identical decision surfaces, and are therefore identical for the purposes of classification. To build a summary statistic, we average over all possible values of $x$ and $x'$, weighted by the signal and background distributions, yielding the ADO:
\begin{equation}
	\label{eq:ADO}
	\ADO[f,g]  = \int \mathrm{d}x\, \mathrm{d}x'\, p_{\text{sig}}(x) \, p_{\text{bkg}} (x') \, \textrm{DO}[f,g](x,x').
\end{equation}
This evaluates to 1 if the decision functions make the same relative classification decision for every pair, to 0 if the functions make the opposite classification for every pair, and to $\frac{1}{2}$ if there is no consistency in their orderings. Since a decision function can be trivially inverted, the cases of $\ADO=0$ and $\ADO=1$ have the same meaning, so we map $\ADO \rightarrow 1- \ADO$ whenever it is less than $\frac{1}{2}$. The ADO has a similar philosophy to Kendall's rank correlation coefficient~\cite{ktau}, with the key difference that we are comparing inputs drawn from separate signal and background distributions.

To gain intuition for the ADO, note that it has a very similar structure to the AUC in \Eq{eq:AUC}. The AUC is the probability that a single decision function orders objects correctly relative to ground truth. The ADO is the probability that two decision functions order objects in the same way, even if incorrectly. In the case that $f(x) = p_{\text{sig}}(x) / p_{\text{bkg}}(x)$ is the likelihood ratio, then $f(x)$ is an optimal classifier by the Neyman--Pearson lemma, so an $\ADO = 1$ implies that $g(x)$ defines the same optimal decision boundaries as $f(x)$. In most ML applications, one is trying to maximize the AUC or other similar metric of absolute classification performance. Our guided strategies, by contrast, aim to maximize the ADO relative to an already trained ML tool.

There are other similarity metrics that one could use, but they are not as easy to interpret in terms of classification decisions. One way to capture information similarity is to use mutual information, or more appropriate to binary classification, mutual information with the truth~\cite{Larkoski:2014pca}. For our guided strategies, though, we are less interested in whether two decision functions have the same quantity of information available to a classification task, and more interested in quantifying the degree to which two decision functions make classification decisions in the same way. Even if $f(x)$ and $g(x)$ contain a high level of mutual information, they do not necessarily define the same decision boundaries. This is especially important to keep in mind given the flexibility of deep networks, which allow the same discrimination power to be encoded in many informationally equivalent ways. Beyond the ADO, there are other summary statistics one could use based on the raw DO information, and we leave a study of those alternatives to future work.

\subsection{Black-Box Guided Search Strategy}
\label{subsec:strategy_blackbox}
\begin{figure*}[ht]
	\centering
	\includegraphics[width=0.95\textwidth]{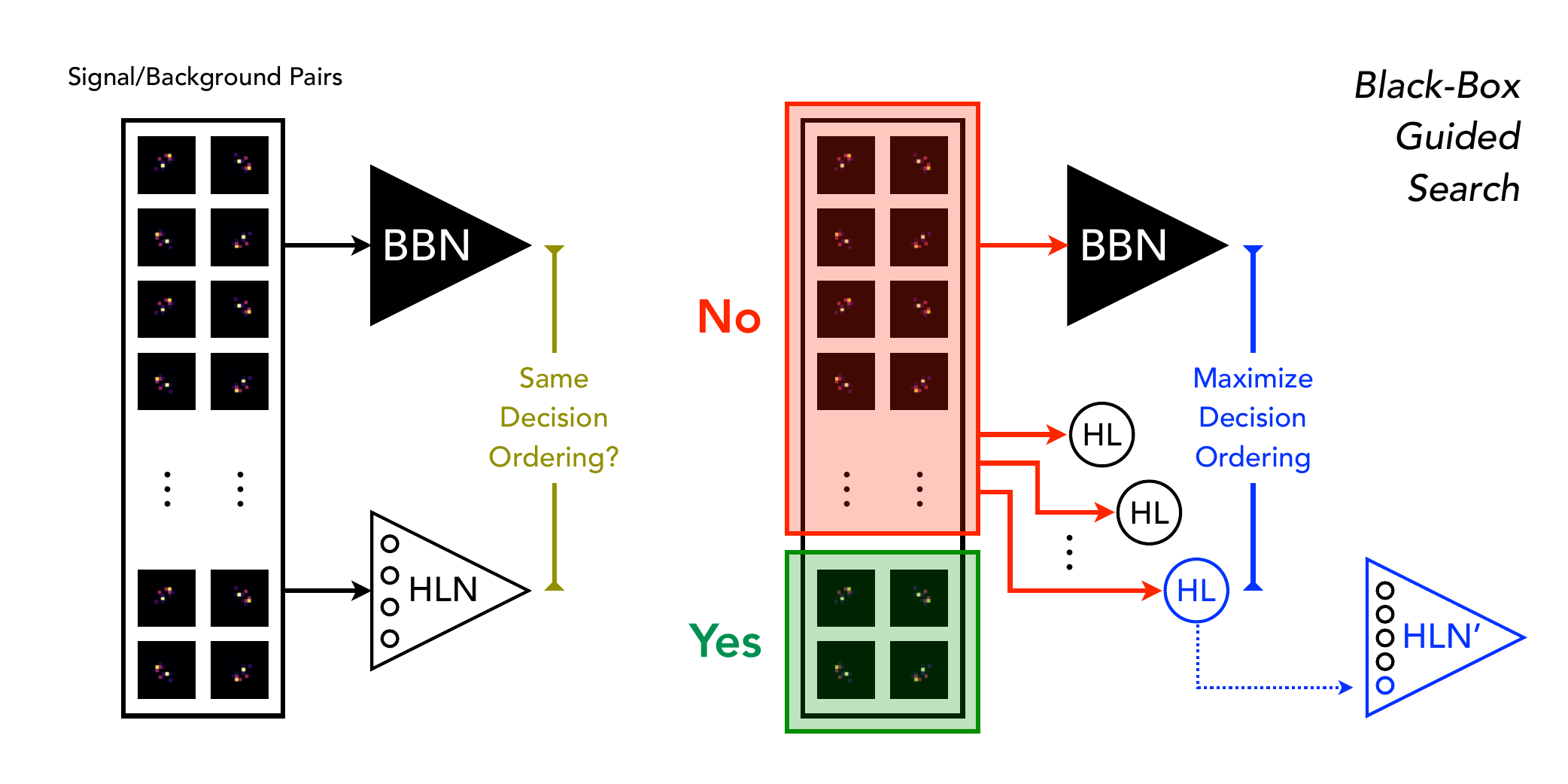}
	\caption{Schematic of the black-box guided search in \Sec{subsec:strategy_blackbox}.
	In each iteration of this strategy, the relative decision ordering of signal/background pairs between the fixed black-box network (BBN, black triangle) and a trainable network of HL observables (HLN, white triangle) is used to identify the subset (red box) in which pairs are differently ordered. From a large space of HL observables (circles), the one with the largest ADO in the misordered space (blue circle) is selected for the next iteration. The schematic above corresponds to the $n=4$ iteration. Note that the BBN is not retrained in each iteration, but the network of HL observables is.}
	\label{fig:tech_bb}
\end{figure*}

The idea behind the black-box guided strategy is shown in \Fig{fig:tech_bb}, where the goal is to find HL observables that maximize the ADO relative to an already trained ML tool. We denote the reference black-box network as ``$\BBN$'' (with apologies to cosmologists), and it will typically be some kind of deep network acting on the low-level  inputs. Starting from a large set $\mathcal{S}$ of human-defined HL observables motivated by physics considerations, our aim is to build a high-level network ($\HLN$) with the same decision surfaces as the $\BBN$. The initial step ($n=0$) in the black-box guided approach is to identify the observable $\HL_1$ that has the largest ADO with the $\BBN$:
\begin{equation}
	\label{eq:HL1_def}
	\HL_1 = \argmax_{\HL \in \mathcal{S}} \ADO[\BBN, \HL]_{X_{\rm all}}.
\end{equation}
Here, the $X_{\rm all}$ subscript indicates that we are using the full set of signal/background training pairs $(x,x')$ when computing the ADO. The observable $\HL_1$ is therefore the physics-motivated observable in the set $\mathcal{S}$ that best approximates the decision surfaces of the black box.

In the next step ($n=1$), we focus our search for HL observables in regions of the feature space where the black-box network disagrees with our current set of HL observables, by isolating the subset of signal/background pairs $X_1$ that are ordered differently by the $\BBN$ and $\HL_1$:
\begin{equation}
	X_1 = \Big\{ (x,x') \, \Big| \, \DO[\BBN, \HL_1](x,x') = 0 \Big\}.
\end{equation}
We then identify the observable $\HL_2$ that has the largest ADO with the $\BBN$ when restricted to the $X_1$ subset:
\begin{equation}
	\label{eq:HL2_def}
	\HL_2 = \argmax_{\HL \in \mathcal{S}} \ADO[\BBN, \HL]_{X_1}.
\end{equation}

For each subsequent step $n > 1$, we combine the HL observables already identified in the previous steps into a joint network:
\begin{equation}
	\label{eq:HLCn_def}
	\HLN_{n} = \NN[\HL_1, \ldots, \HL_{n}],
\end{equation}
where $\NN$ indicates a neural network trained on the full signal/background training set with just the $n$ HL observables as inputs. From this joint $\HLN$, the misordered subset $X_{n}$ is defined via
\begin{equation}
	\label{eq:Xndef}
	X_{n} = \Big\{ (x,x') \, \Big| \, \DO[\BBN,\HLN_n](x,x') = 0 \Big\}.
\end{equation}
Because a new $\HLN$ is trained in each iteration, $X_{n}$ may not be a strict subset of $X_{n-1}$. The next observable $\HL_{n+1}$ is determined via:
\begin{equation}
	\label{eq:HLn1_def}
	\HL_{n+1} = \argmax_{\HL \in \mathcal{S}} \ADO\big[\BBN, \HL \big]_{X_{n}}.
\end{equation}
Note that the same black-box network is used in each iteration, but the changing subset $X_{n}$ means that different decision surface information is tested at each step. These steps are repeated until $\ADO[\BBN, \HLN_{n+1}]$ gets as close to 1 as desired.

Isolating the differently-classified pairs in \Eq{eq:Xndef} is similar in spirit to the boosting step of BDTs~\cite{10.1023/A:1022648800760,FREUND1995256}. This approach focuses attention only on the subspace of pairs where the $\BBN$ disagrees with the current set of HL observables, allowing us to identify new HL observables that make signal-background ordering decisions most similar to the BBN in that subspace. It is worth emphasizing that the ADO, or some other metric for network decision similarity, is essential for this approach to work.

Later in \Sec{subsec:iteratively_truthlabels}, we will compare this black-box guided approach to a label guided approach. Instead of using the ADO, the label guided approach uses the AUC with respect to ground truth information. It is straightforward to understand why the ADO is superior to the AUC for guiding purposes. To the extent that the $\BBN$ is well trained, it represents a good approximation to the Neyman--Pearson optimal classifier. Achieving the correct DO relative to the optimal classifier for every signal/background training pair is the best one could ever hope to do. Therefore, if the black-box guiding strategy is working correctly, then the subsets $X_{n}$ will get smaller and smaller until almost all signal/background pairs have been correctly ordered relative to the $\BBN$.

By contrast, the AUC captures DO relative to truth labels. Unless the $\BBN$ is able to achieve $\AUC = 1$, there will inevitably be signal/background pairs that are incorrectly ordered even by the theoretically optimal classifier. Instead of getting smaller and smaller, the subsets $X_{n}$ will stall at the set of signal/background pairs that can never be ordered correctly. This in turn means that the classification performance of $\HLN_{n}$ will stall well below the theoretical maximum in the label guided approach. That is why we advocate for the selection of HL observables to be guided by the ADO, since then the classification performance of the $\HLN_{n}$ will eventually match that of the $\BBN$, as desired.

As with any ``greedy algorithm'', our black-box guided strategy cannot identify situations where two HL observables could be combined simultaneously to match the $\BBN$ decision surfaces. This means that we might miss sets of observables that are individually poor classifiers but perform well jointly. If the goal were to just to maximize performance, this would be an undesirable feature. In the context of mapping a black-box ML strategy to a physically-interpretable space, though, we are indeed looking for individual observables with high information content relevant for classification, so this greedy strategy is the one most likely to yield physical insight. 

\section{A Case Study in Jet Substructure}
\label{sec:casestudy}
We now apply the  technique introduced in \Sec{sec:strategy} to a specific case study involving jet classification at the LHC. In this section, we review boosted $W$ boson classification using jet substructure and highlight the elements of \rref{Baldi:2016fql} that will be used in our case study. We then introduce the EFPs~\cite{Komiske:2017aww} as our set of HL physics-motivated observables. Details about our NN architectures and training parameters are provided in \App{app:casestudy_nn}.

\subsection{Boosted Boson Classification}
\label{subsec:casestudy_boosted}
Massive objects produced at the LHC often have enough transverse momentum that their decay products become collimated. For an object with a hadronic decay mode, such as a $W$ boson decaying to a quark-antiquark pair ($W\rightarrow q\bar{q}'$), the resulting jet in the detector consists of two clusters of energy, one from each of the fragmenting quarks. The substructure of these jets is distinct from those that arise from the fragmentation of a single hard quark or gluon. Identification of jets with nontrivial substructure has become an essential tool for probing the nature of collisions at the LHC~\cite{Seymour:1991cb,Seymour:1993mx,Butterworth:2002tt,Butterworth:2007ke,Butterworth:2008iy,Abdesselam:2010pt,Altheimer:2012mn,Shelton:2013an,Altheimer:2013yza,Adams:2015hiv,Larkoski:2017jix,Asquith:2018igt,Marzani:2019hun}

There are many different ways to represent the information in a jet. At the most fundamental level, a jet is variable-length collection of four-vectors with associated particle properties, motivating set-based ML tools~\cite{Komiske:2018cqr,Qu:2019gqs,Moreno:2019bmu,Mikuni:2020wpr,Bogatskiy:2020tje,Shlomi:2020ufi}. Another popular approach is to describe a jet as a grid of calorimeter cells with energy depositions, giving rise to a ``jet image''~\cite{Cogan:2014oua,deOliveira:2015xxd}. In any of these low-level representations, the jet data is high dimensional. This motivates the development of HL observables that intelligently summarize the low-level  information to reduce the effective dimensionality of the task. Physicists have engineered numerous HL observables tasks that incorporate domain knowledge about jet formation (see
\rrefs{Butterworth:2008iy,Kaplan:2008ie,Almeida:2008yp,Ellis:2009me,Plehn:2010st,Cui:2010km,Thaler:2010tr,Thaler:2011gf,Ellis:2012sn,Dasgupta:2013ihk,Larkoski:2013eya,Larkoski:2014wba,Aad:2014haa,Larkoski:2014gra,Buckley:2020kdp} for an incomplete list). Typical usage is to apply cuts on one or more of these HL observables, or to combine several of them using a shallow ML classifier.

In the context of jet classification, ML tools based on low-level  inputs have outperformed traditional strategies based on HL observables~\cite{Kasieczka:2019dbj}. Of course, the HL observables themselves are just functions of the low-level  inputs, so it should be possible to find a large enough set of physics-motivated HL observables that can match the performance of these ML classifiers~\cite{Datta:2017rhs,Moore:2018lsr,Aguilar-Saavedra:2020sxp}. This is indeed the intuition behind the guided strategy in \Sec{sec:strategy}, where the goal is to leverage a black-box ML method to identify the most effective HL observables.

Our case study is based on the same datasets as \rref{Baldi:2016fql}. These datasets correspond to $\sqrt{s}=14\, \mbox{TeV}$ proton-proton collision, where hard scattering and resonance decay were generated using \textsc{MadGraph 5} v2.2.3~\cite{Alwall:2011uj}, showering and hadronization were generated with \textsc{Pythia} v6.426~\cite{Sjostrand:2006za}, and the response of the detectors was simulated with \textsc{Delphes} v3.2.0~\cite{deFavereau:2013fsa}. The boosted $W$ signal process is diboson production ($pp \to W^+ W^-$), which yields two fat jets each with 2-prong substructure. The background process is QCD dijet production ($pp \to qq, qg, gg$), which typically yields 1-prong jets. These samples do not include contamination from pileup (multiple proton-proton collisions per beam crossing). Jets are clustered using the anti-$k_{\textrm{t}}$ algorithm~\cite{Cacciari:2008gp} with radius parameter $R = 1.2$, using \textsc{FastJet 3.1.2}~\cite{Cacciari:2011ma}. The dataset contains $5 \times 10^{6}$ events, split equally between signal and background. Following the approach in \rref{Baldi:2016fql}, each jet is pixelated into a $32\times32$ grid in the rapidity-azimuth plane, and a jet image is formed from the transverse momentum ($p_{\textrm{T}}$) deposits in each cell. The jet image is then trimmed \cite{Krohn:2009th}, where subjets of radius $R_{\rm sub}=0.2$ are discarded if their $p_{\textrm{T}}$ is less than 3\% of the original jet. The final jet selection takes jets with trimmed momentum $p^{\rm trim}_{\textrm{T}} \in [300,400]$ GeV within the rapidity range $\left|\eta\right| < 5.0$. While important jet information is lost by pixelation and trimming, we include these steps in our analysis in order to perform an apples-to-apples comparison to \rref{Baldi:2016fql}.

\begin{table}[t]
	\centering
	\begin{tabular}{lcc}
		\hline \hline
		Observable         & $\AUC$             & $\ADO[\CNN,\text{Obs.}]$ \\
		\hline \hline
		$M_{\text{jet}}$   & $0.898 \pm 0.004$  & 0.807 \\
		$C_2^{\beta=1}$    & $0.660 \pm 0.006$  & 0.584 \\
		$C_2^{\beta=2}$    & $0.604 \pm 0.007$  & 0.548 \\
		$D_2^{\beta=1}$    & $0.790 \pm 0.005$  & 0.743 \\
		$D_2^{\beta=2}$    & $0.807 \pm 0.005$  & 0.762 \\
		$\tau_2^{\beta=1}$ & $0.662 \pm 0.006$  & 0.600 \\
		\hline
		6HL                & $0.9504 \pm 0.0002$  & 0.971 \\
		CNN                & $0.9531 \pm 0.0002$  & 1.000 \\
		488HL   & $0.9535 \pm 0.0002$ & 0.978\\
		\hline
		$7\HL_{\text{black-box}}$ & $0.9528 \pm 0.0003$   & 0.971 \\
		\hline\hline
	\end{tabular}
	\caption{Classification performance of the six HL observables studied in \rref{Baldi:2016fql}, as well as a 6HL joint classifier.
	The six HL observables face a small but significant performance gap compared to the benchmark CNN. As discussed later in \Sec{subsec:supplementing_blackbox}, this performance gap is bridged by a seventh feature discovered using our black-box guided strategy.  The 488HL case involving a large set of EFPs is discussed at the end of \Sec{subsec:iteratively_bruteforce}.  The uncertainty on the AUC is computed from 1 standard deviation of 10-fold cross-validation. The decision similarity (ADO) to the benchmark CNN is also shown. Details of the NN architectures are provided in \App{app:casestudy_nn}.}
	\label{tab:AUC6HL}
\end{table}

The trimmed jet's constituents are used to compute six HL jet substructure observables:  the trimmed jet mass ($M_{\textrm{jet}}$), four ratios of energy correlation functions ($C^{\beta=1}_{2}$, $C^{\beta=2}_{2}$, $D^{\beta=1}_{2}$, $D^{\beta=2}_{2}$)~\cite{Larkoski:2013eya,Larkoski:2014gra}, and
the $N$-subjettiness ratio ($\tau^{\beta=1}_{21}$)~\cite{Thaler:2010tr,Thaler:2011gf}. These observables are well-established in the context of boosted $W$ classification, including studies at ATLAS~\cite{Aad:2015rpa,Aaboud:2018psm} and CMS~\cite{Khachatryan:2014vla}. The $W$ boson classification performance of these six HL observables is summarized in \Tab{tab:AUC6HL}. The trimmed jet mass is the most powerful single observable, since the 80.4 GeV mass peak is a characteristic feature of boosted $W$ bosons.

\begin{figure}[t]
	\centering
	\includegraphics[width=0.48\textwidth]{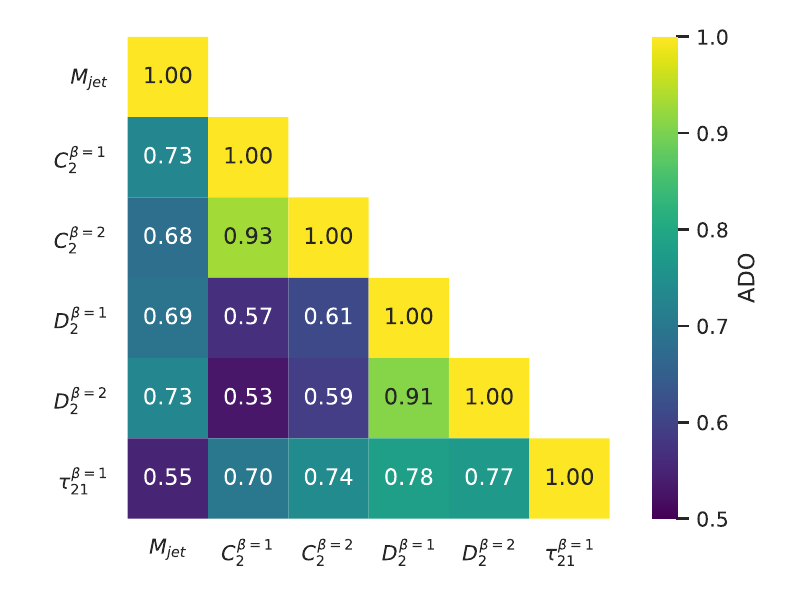}
	\caption{Similarity of the classification decisions between the six HL observables, as quantified by the ADO. A value of $\ADO=1$ indicates identical decision ordering for all signal/background pairs, while $\ADO=\frac{1}{2}$ corresponds to no similarity. In this way, the ADO has a similar interpretation to the AUC, but with respect to classification decisions instead of ground truth.}
	\label{fig:HL_vs_HL_ADO2D}
\end{figure}

We can use the ADO from \Eq{eq:ADO} to gain additional insight into these six HL observables. In \Fig{fig:HL_vs_HL_ADO2D}, we assess the pairwise ADO between each of the HL observables considered. The observable pairs that make the most similar decisions (i.e.\ $\ADO \rightarrow 1$) are $C^{\beta=1}_{2}$ with $C^{\beta=2}_{2}$ and $D^{\beta=1}_{2}$ with $D^{\beta=2}_{2}$. This is expected since these observables have relatively similar structures except for the choice of $\beta$ coefficient, which controls the weighting of angular information within the jets. These pairs also have similar AUC values, as seen in \Tab{tab:AUC6HL}, since pairs that make common classification decisions should exhibit similar classification power. Comparing the AUC and ADO values provides a more detailed picture about the degree of correlation in classification.

\begin{figure}[t]
	\centering
	\includegraphics[width=0.48\textwidth]{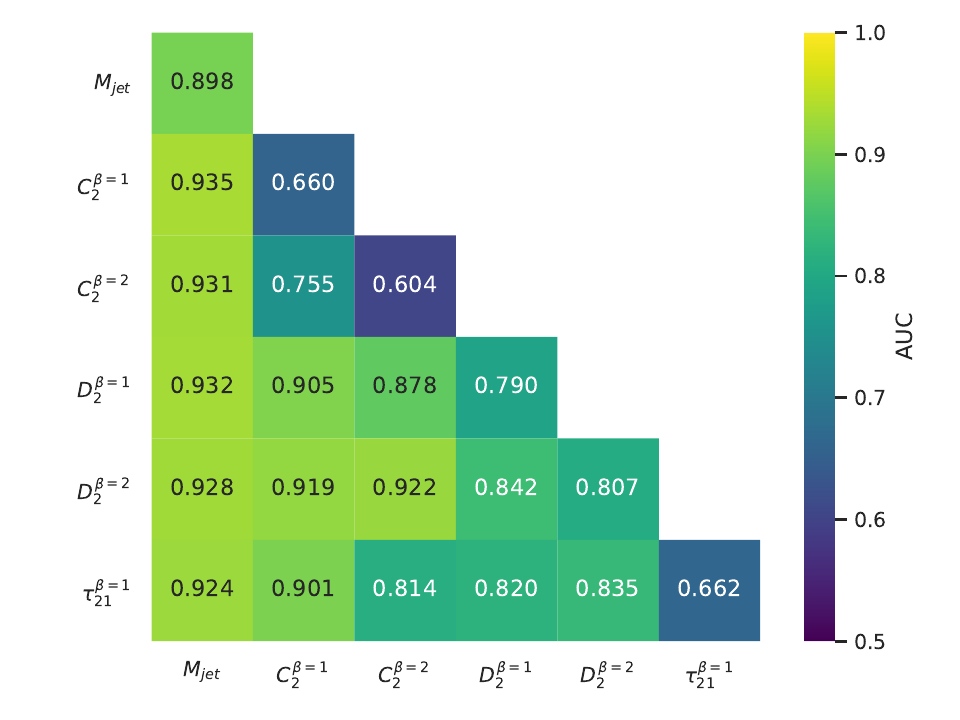}
	\caption{Classification performance of the six HL observables in \Tab{tab:AUC6HL} on the diagonal entries, along with AUC values for pairwise classification between coupled HL inputs on the off-diagonal entries.
	}
	\label{fig:pairwiseAUC6HL}
\end{figure}

The observable pairs that make the least similar decisions (i.e.\ $\ADO \rightarrow \frac{1}{2}$) are $M_{\text{jet}}$ with $\tau^{\beta=1}_{21}$ and $C^{\beta}_{2}$ with $D^{\beta}_{2}$. For $M_{\text{jet}}$ versus $\tau^{\beta=1}_{21}$, this is expected since $N$-subjettiness probes the degree of prong-like collimation, whereas mass is sensitive to the energies of the prongs and their relative angles. For $C^{\beta}_{2}$ versus $D^{\beta}_{2}$, this is expected since they have different scalings under boosts along the jet direction~\cite{Larkoski:2014gra}. Pairs that make dissimilar decisions can often be combined into more powerful joint classifiers. This is shown in \Fig{fig:pairwiseAUC6HL}, where we consider the pairwise classifiers $\NN[\HL_i,\HL_j]$, where the details of the NN parameters are presented in \App{app:dnn_hl}. More comprehensive studies of these six jet substructure observables can be found in \rref{Adams:2015hiv}.

While these six engineered HL features are powerful jet substructure discriminants, they do not capture the full information relevant for $W$ boson tagging. Viewing the calorimeter cells as pixels of a two-dimensional image, we can try to enhance the discrimination power using computer vision techniques~\cite{Cogan:2014oua,deOliveira:2015xxd,Almeida:2015jua,Baldi:2016fql,Barnard:2016qma,Komiske:2016rsd,ATL-PHYS-PUB-2017-017,Kasieczka:2017nvn,Macaluso:2018tck}. Indeed, \rref{Baldi:2016fql} demonstrated that a deeply connected CNN using the low-level  jet image inputs yielded better classification performance than the six HL observables combined with a BDT. The performance gain was modest though persistent, making it an excellent benchmark problem for studying the interpretation of ML strategies. We repeat this exercise in \Tab{tab:AUC6HL}, now using the NN parameters in \App{app:casestudy_nn} for the CNN and for the 6HL combination:
\begin{equation}
	6\HL \equiv \NN\big[M_{\text{jet}}, C_2^{\beta=1}, C_2^{\beta=2},D_2^{\beta=1},D_2^{\beta=2},\tau_2^{\beta=1}\big],
\end{equation}
where we find a $0.0027 \pm 0.0003$ gap in AUC, as seen in \Tab{tab:AUC6HL}. Using our guided strategy, we seek to understand the nature of this performance gap, and whether the CNN has found a strategy similar to the existing HL observables or something distinct. Does the gap indicate a mild optimization of the same basic HL ideas, or does it hint at the existence of a new HL observable not appearing previously in the jet substructure literature?

\subsection{Energy Flow Polynomials}
\label{subsec:casestudy_efps}
In order to map the CNN from \rref{Baldi:2016fql} to a human-readable space, we first define a suitable set of physics-motivated HL observables for use in the guided strategies. This requires domain knowledge about the underlying physics as well as intuition for the kinds of information that might be missing from existing HL observables. This also requires identifying HL observables that are likely to work well as classifiers individually, since the black-box guided strategy in \Sec{subsec:strategy_blackbox} is based on finding a single observable that maximizes the ADO in each step.

Our set of HL observables is based on the EFPs~\cite{Komiske:2017aww}. The EFPs are a large (formally infinite) set of parametrized engineered functions, inspired by previous work on energy correlation functions~\cite{Banfi:2004yd,GurAri:2011vx,Jankowiak:2011qa,Larkoski:2013eya,Larkoski:2014gra,Moult:2016cvt}. In the jet image representation, they are defined in terms of the momentum fraction $z_a$ of calorimeter cell $a$, as well as the pairwise angular distance $\theta_{ab}$ between cells $a$ and $b$. The EFPs are built in increasing levels of complexity, from simple sums over single cells to many higher-order combinations of momentum and pair-wise angles. An EFP can be represented as a multigraph where:
\begin{align}
	\text{each node} \quad          & \Longrightarrow \quad \sum_{a = 1}^N z_a, \label{eq:EFP_node}          \\
	\text{each $k$-fold edge} \quad & \Longrightarrow \quad \left(\theta_{ab}\right)^k \label{eq:EFP_edge} .
\end{align}
As one example, we have:
\begin{equation}
         \label{eq:g81}
	\ndk{0.045}{4}{7}{20} = \sum_{a=1}^N \sum_{b=1}^N \sum_{c=1}^N \sum_{d=1}^N z_a z_b z_c z_d \theta_{ab}^3 \theta_{bc} \theta_{ac} \theta_{ad}^2 .
\end{equation}
From these graph representations, we can express both connected and disconnected graphs. For the purposes of our study, we only consider connected graphs.

The observable corresponding to each graph can be modified with parameters $(\kappa, \beta)$. These parameters determine the specific meaning of $z_a$ and $\theta_{ab}$, where
\begin{align}
	z^{(\kappa)}_a        & = \left(\frac{p_{\textrm{T}a}}{\sum_b p_{\textrm{T}b}} \right)^\kappa, \label{eq:EFP_z} \\
	\theta^{(\beta)}_{ab} & = \left(\Delta \eta_{ab}^2 + \Delta \phi_{ab}^2 \right)^{\beta/2}. \label{eq:EFP_theta}
\end{align}
Here, $p_{\textrm{T}a}$ is the transverse momentum of cell $a$, and $\eta_{ab}$ ($\phi_{ab}$) is the pseudorapidity (azimuth) difference between cells $a$ and $b$. The original IRC-safe EFPs require $\kappa = 1$, however we include examples with $\kappa \not= 1$ to explore a broader space of HL observables, motivated by \rrefs{Pandolfi:2012ima,Chatrchyan:2012sn,Larkoski:2014pca,Gras:2017jty}.\footnote{We thank Patrick Komiske and Eric Metodiev for discussions related to the normalization of the $\kappa \not= 1$ EFPs.} Note that $\kappa > 0$ generically corresponds to IR-safe but C-unsafe observables. We intentionally include zero and negative values of $\kappa$ to explore both IR-unsafe and C-unsafe observables as well.\footnote{For $\kappa < 0$, empty cells are omitted from the sum in \Eq{eq:EFP_node}.  This is not as discontinuous as one might naively think due to the pixelation and trimming steps.} For our study, we use the \textsc{EnergyFlow} python package \cite{efpython}  to translate jet-image $p_\textrm{T}$, $\eta$, and $\phi$ information to EFPs with varying graphs and choices of $\kappa$ and $\beta$.

For our guided search, we consider all combinations of $(\kappa, \beta)$ where $\kappa = [-1, 0, \frac{1}{2}, 1, 2]$ and $\beta=[\frac{1}{2}, 1, 2]$. Each of the 15 combinations of $(\kappa, \beta)$ are applied to the complete set of connected graphs with degree (i.e.\ number of edges) $d \leq 7$ along with all connected graphs with degree $d\leq 8$ and chromatic number $c = 4$ (to be defined in \Sec{subsec:supplementing_interpretation} below), which comes to 509 in total. This yields a space of 7,635 HL observables to search from. Note that $\beta$ in \Eq{eq:EFP_theta} can sometimes be traded for $k$ in \Eq{eq:EFP_edge}; we remove degenerate graphs from our space, reducing our pool of unique observables to 7,545.

It is important to emphasize that, although the EFP space constitutes a formally complete basis for (IRC-safe) jet classification, we are primarily concerned with the pragmatics of isolating individual observables that can map out the CNN behavior. The ideal case is that the CNN strategy maps to a single EFP, indicating that it can be expressed compactly in terms that can be easily interpreted by humans. Failing that, though, it is still of considerable value if a similar mapping can be made using a small collection of observables~\cite{Datta:2017rhs,Moore:2018lsr,Aguilar-Saavedra:2020sxp}. This would still provide a significantly more physically meaningful interpretation of the data and a marked reduction in data complexity and dimensionality.

Beyond this specific benchmark example, if one is unable to map the CNN strategy into a small number of EFPs, this could mean one of two things. First, it could mean that the CNN strategy simply does not operate in this HL space, requiring us to revisit the assumption that the HL space was sufficiently complete to capture the essential information for jet classification. Second, it could mean that the CNN strategy is still encodable in terms of these HL observables, but in a more complex combination. As an example of this second possibility, consider the $C_2$~\cite{Larkoski:2013eya} and $D_2$~\cite{Larkoski:2014gra} observables discussed in \Sec{subsec:casestudy_boosted}. These can be written as EFPs with $\kappa = 1$:
\begin{align}
	C^{(\beta)}_2 & = \ndk{0.045}{3}{3}{1} \bigg/ \bigg(\ndk{0.045}{2}{1}{0}\bigg)^2, \label{eq:C2EFP} \\
	D^{(\beta)}_2 & = \ndk{0.045}{3}{3}{1} \bigg/ \bigg(\ndk{0.045}{2}{1}{0}\bigg)^3, \label{eq:D2EFP}
\end{align}
where the graphs corresponds to:
\begin{align}
	\ndk{0.045}{3}{3}{1} & = \sum^N_{a=1} \sum^N_{b=1} \sum^N_{c=1} z_a z_b z_c \theta^{(\beta)}_{ab} \theta^{(\beta)}_{bc} \theta^{(\beta)}_{ca}, \label{eq:complete3} \\
	\ndk{0.045}{2}{1}{0} & = \sum^N_{a=1} \sum^N_{b=1} z_a z_b \theta^{(\beta)}_{ab} \label{eq:complete2}.
\end{align}
The guided strategies, however, would not necessarily be able to identify these ratio combinations unless they were defined ahead of time.\footnote{It might be interesting to combine the guided strategy with some kind of symbolic regression to find interesting combinations~\cite{Udrescu:2019mnk}.  If this symbolic regression allowed for index contraction, then one could use the energy flow moments~\cite{Komiske:2019asc} to more efficiently search the space of $\beta = 2$ EFPs.}
Therefore, whether or not the guided mapping is effective, one learns something about the nature of the physics problem either way.

\section{Supplementing Existing Observables}
\label{sec:supplementing}
In this section, we demonstrate the success of the mapping strategy from \Sec{sec:strategy} in searching for an additional HL observable in the context of boosted $W$ classification.

From \Tab{tab:AUC6HL}, we saw that the difference in classification power between a CNN acting on jet images and an NN combination of six HL observables is relatively small, but genuine and statistically significant. As such, it is interesting to ask whether this existing set of six HL observables could be supplemented by a new single HL observable which has not yet been considered by human physicists. We employ our black-box  guiding strategy to find such a seventh HL observable, which closes the performance gap previously identified in \rref{Baldi:2016fql}.

\subsection{Black-Box Guiding}
\label{subsec:supplementing_blackbox}

\begin{table*}
\begin{tabular}{c|cccc|cc|cc}
	\hline \hline
	Rank & EFP & $\kappa$ & $\beta$ & Chrom \# & $\text{ADO}[\text{EFP}, \text{CNN}]_{X_{6}}$ & AUC[EFP] & $\text{ADO}[\text{6HL + EFP},\text{CNN}]_{X_{\rm all}}$ & $\text{AUC}[\textrm{6HL + EFP}]$\\ \hline \hline
	&&&&&&& \\[-0.6em]
1 & $\tndk{4}{4}{3}$ & 2 & $\frac{1}{2}$ & 3 & 0.6207 & 0.8031 & 0.9714 & $0.9528 \pm 0.0003$ \\
2 & $\tndk{6}{7}{99}$ & 2 & $\frac{1}{2}$ & 3 & 0.6205 & 0.8203 & 0.9714 & 0.9524 \\
3 & $\tndk{1}{0}{0}$ & 0 & -- & 1 & 0.6205 & 0.6737 & 0.9715 & 0.9525 \\
4 & $\tndk{6}{7}{96}$ & 2 & $\frac{1}{2}$ & 3 & 0.6199 & 0.8301 & 0.9715 & 0.9527 \\
5 & $\tndk{6}{7}{95}$ & 2 & $\frac{1}{2}$ & 3 & 0.6197 & 0.8290 & 0.9714 & 0.9527 \\
6 & $\tndk{6}{6}{21}$ & 2 & $\frac{1}{2}$ & 3 & 0.6196 & 0.8251 & 0.9715 & 0.9522 \\
7 & $\tndk{2}{1}{0}$ & 0 & $\frac{1}{2}$ & 2 & 0.6187 & 0.7511 & 0.9715 & 0.9526 \\
8 & $\tndk{5}{6}{18}$ & 2 & $\frac{1}{2}$ & 3 & 0.6184 & 0.8257 & 0.9712 & 0.9527 \\
9 & $\tndk{4}{6}{19}$ & 2 & $\frac{1}{2}$ & 3 & 0.6182 & 0.8090 & 0.9714 & 0.9527 \\
10 & $\tndk{6}{7}{67}$ & 2 & $\frac{1}{2}$ & 3 & 0.6180 & 0.8314 & 0.9714 & 0.9526 \\
\hline
60 & $\tndk{2}{1}{0}$ & 0 & 1 & 2 & 0.6163 & 0.7194 & 0.9715 & 0.9525 \\
341 & $\tndk{4}{7}{36}$ & $-1$ & $\frac{1}{2}$ & 4 & 0.6142 & 0.6286 & 0.9714 & 0.9509 \\
589 & $\tndk{2}{1}{0}$ & 0 & 2 & 2 & 0.6109 & 0.7579 & 0.9714 & 0.9523 \\
3106 & $\tndk{1}{0}{0}$ & $-1$ & --  & 1 & 0.5891 & 0.5882 & 0.9714 & 0.9510 \\
3519 & $\tndk{8}{7}{0}$ & $\frac{1}{2}$ & $\frac{1}{2}$ & 2 & 0.5664 & 0.7698 & 0.9715 & 0.9524 \\
3521 & $\tndk{1}{0}{0}$ & $\frac{1}{2}$ & -- & 1 & 0.5663 & 0.7093 & 0.9714 & 0.9522 \\
5531 & $\tndk{8}{7}{0}$ & 1 & 2 & 1 & 0.5290 & 0.7454 & 0.9714 & 0.9507 \\
5554 & $\tndk{2}{1}{0}$ & 1 & $\frac{1}{2}$ & 2 & 0.5279 & 0.8210 & 0.9713 & 0.9505 \\
5610 & $\tndk{1}{0}{0}$ & 2 & -- & 1 & 0.5245 & 0.7117 & 0.9714 & 0.9507 \\
\hline
5657 & $\tndk{3}{3}{1}$ & 1 & 1 & 3 & 0.5224 & 0.8257 & 0.9712 & 0.9506 \\
5793 & $\tndk{2}{1}{0}$ & 1 & 1 & 2 & 0.5191 & 0.8640 & 0.9714 & 0.9505 \\
6052 & $\tndk{3}{3}{1}$ & 1 & 2 & 3 & 0.5153 & 0.8500 & 0.9716 & 0.9504 \\
7438 & $\tndk{2}{1}{0}$ & 1 & 2 & 2 & 0.5011 & 0.8835 & 0.9716 & 0.9506 \\
\hline \hline
\end{tabular}
	\centering
	\caption{A selection of EFPs, sorted by their similarity  with the CNN, evaluated using ADO in the differently-ordered subspace $X_6$. This corresponds to one step in the black-box guiding technique depicted in \Fig{fig:tech_bb}. After the top 10, EFPs are shown if they correspond to a dot graph, appear in the $C_2$/$D_2$ observables from \Eqs{eq:C2EFP}{eq:D2EFP}, or have the highest ADO among graphs with a given value of $\kappa$, $\beta$, or chromatic number.}
	\label{tab:EFPvsBB}
\end{table*}

The first step of the black-box guided strategy from \Sec{subsec:strategy_blackbox} is to identify the subset of signal/background pairs that are differently ordered by the $\CNN$ and the $6\HL$ combination:
\begin{equation}
	X_6 = \Big\{ (x,x') \, \Big| \, \DO\big[\CNN, 6\HL \big](x,x') = 0 \Big\}.
\end{equation}
Though we have $6.25 \times 10^{12}$ signal/background pairs, we construct $X_6$ from a randomly selected subset of $5 \times 10^7$ pairs. We then search for the EFP that has the greatest decision similarity to the $\CNN$ in the $X_6$ subspace, which ideally captures all of the remaining discrepancies between the $\CNN$ and $6\HL$ approaches:
\begin{equation}
	\label{eq:HL7_def}
	\HL^{\text{black-box}}_7 = \argmax_{\HL \in \text{EFPs}} \ADO[\CNN, \HL]_{X_6}.
\end{equation}

The results are shown in the first row of \Tab{tab:EFPvsBB}. The EFP with the largest ADO with the CNN in the $X_6$ subspace is:
\begin{equation}
	\label{eq:g8_again}
	\ndk{0.045}{4}{4}{3} {}^{\left(\kappa = 2, \beta = \tfrac{1}{2}\right)} = \sum_{a=1}^{N} \sum_{b=1}^{N} \sum_{c=1}^{N} \sum_{d=1}^{N} z^{2}_a z^{2}_b z^{2}_c z^{2}_d \sqrt{\theta_{ab} \theta_{bc} \theta_{ac} \theta_{ad}}.
\end{equation}
By itself, \Eq{eq:g8_again} only has an $\AUC$ of 0.8031,  but when used as the seventh feature of an NN that also uses the previously identified 6HL observables,
\begin{equation}
	7\HL_{\text{black-box}} \equiv \NN\left[M_{\text{jet}}, \ldots,\tau_2^{\beta=1}, 	\ndk{0.045}{4}{4}{3} {}^{\left(\kappa = 2, \beta = \tfrac{1}{2}\right)} \right],\\
\end{equation}
it closes the performance gap with the $\CNN$ by achieving $\AUC = 0.9528 \pm 0.0003$, as previewed in \Tab{tab:AUC6HL}. Interestingly, this happens even though the ADO between the $\CNN$ and $7\HL_{\text{black-box}}$ is only $0.971$, implying that they still make inconsistent decisions around 3\% of the time. So while the black box has guided the selection of a new HL observable that closes the AUC performance gap, the remaining ADO gap implies that there is additional information not being captured.

The remaining rows of \Tab{tab:EFPvsBB} show portions of the ranked list of 7,545 EFPs, ordered by their ADO values. Note that the statistical uncertainties on the ADO are large enough that the precise ranking is not so meaningful, though the overall trends are. One striking feature is that many observables have a similar ADO to \Eq{eq:g8_again}, and they often feature $\kappa = 2$ and $\beta = \frac{1}{2}$. Recall that $\kappa = 2$ corresponds to IRC-unsafe EFPs, which suggests that IRC-unsafe information is valuable (though perhaps not uniquely so) for mapping the $\CNN$ strategy. Similarly, the appearance of $\beta = \frac{1}{2}$ suggests the importance of probing small-angle behavior. Other IRC-unsafe EFPs with $\kappa = 0$ and $\kappa = -1$ also perform well, especially the constituent multiplicity appearing third on this list. The best performing IRC-safe $\kappa = 1$ observable appears 5531th on this list and is not able to close the performance gap with the CNN. Specifically, the EFPs in \Eqs{eq:complete3}{eq:complete2} with $\kappa = 1$ have a relatively small ADO in the $X_6$ subspace, never getting above $0.5279$.  This is as one might expect, since these observables already effectively appear in the $C_2$ and $D_2$ combinations. Further discussions of the physics implications are provided below.

\begin{figure}[t]
\centering
	\includegraphics[width=0.48\textwidth]{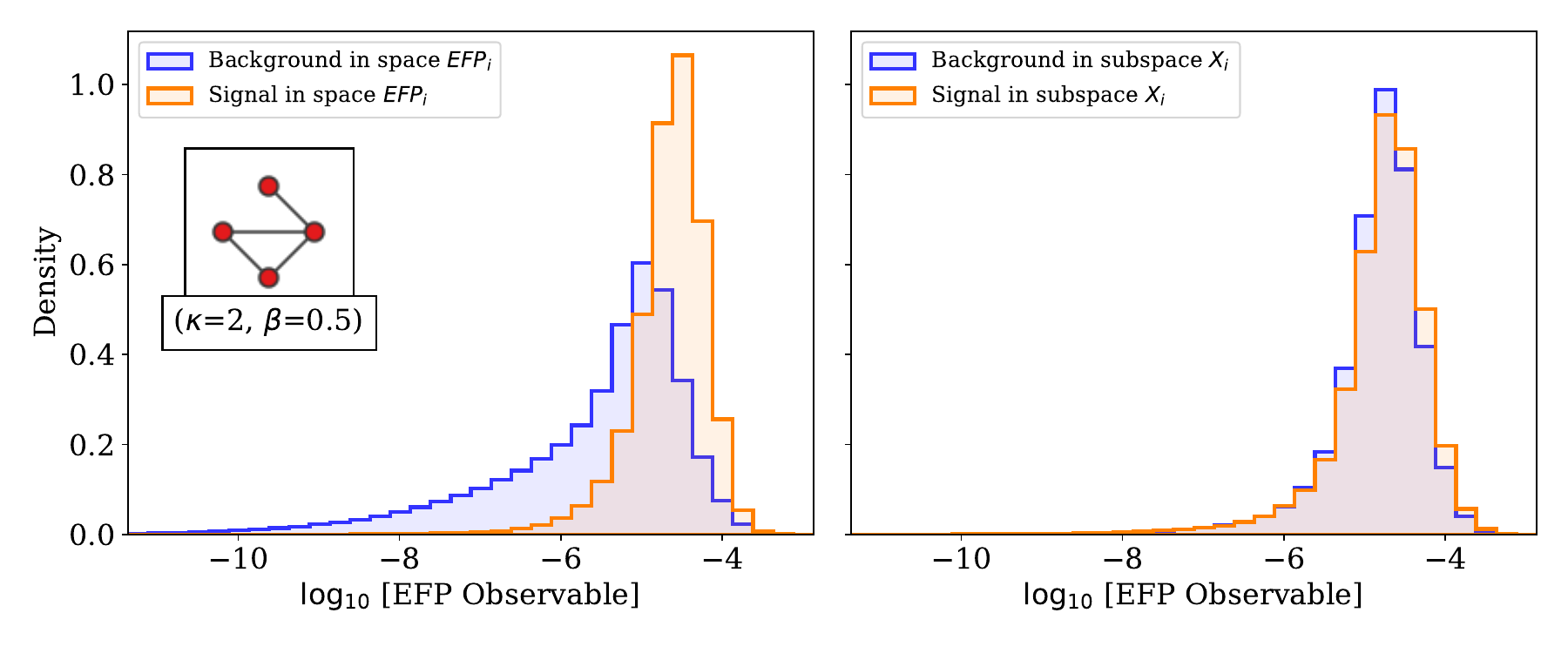}
	\includegraphics[width=0.48\textwidth]{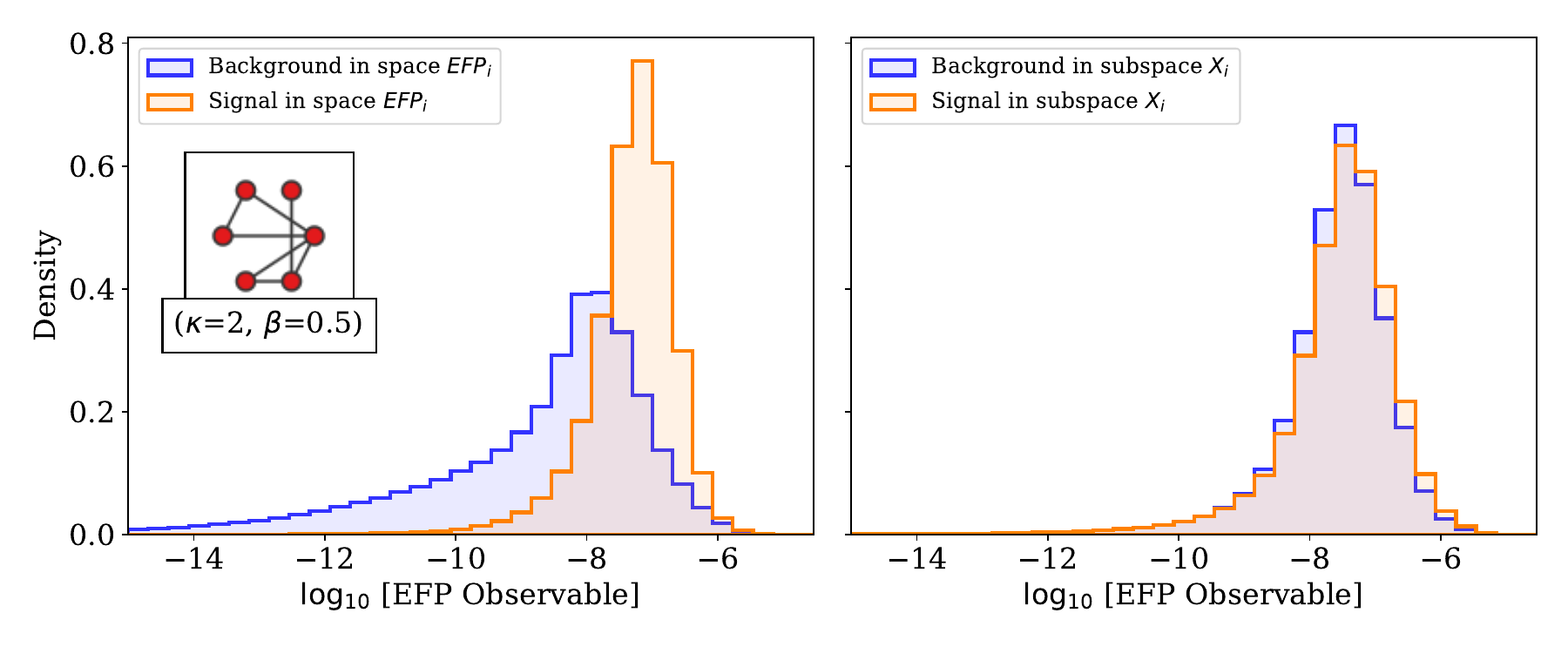}
	\includegraphics[width=0.48\textwidth]{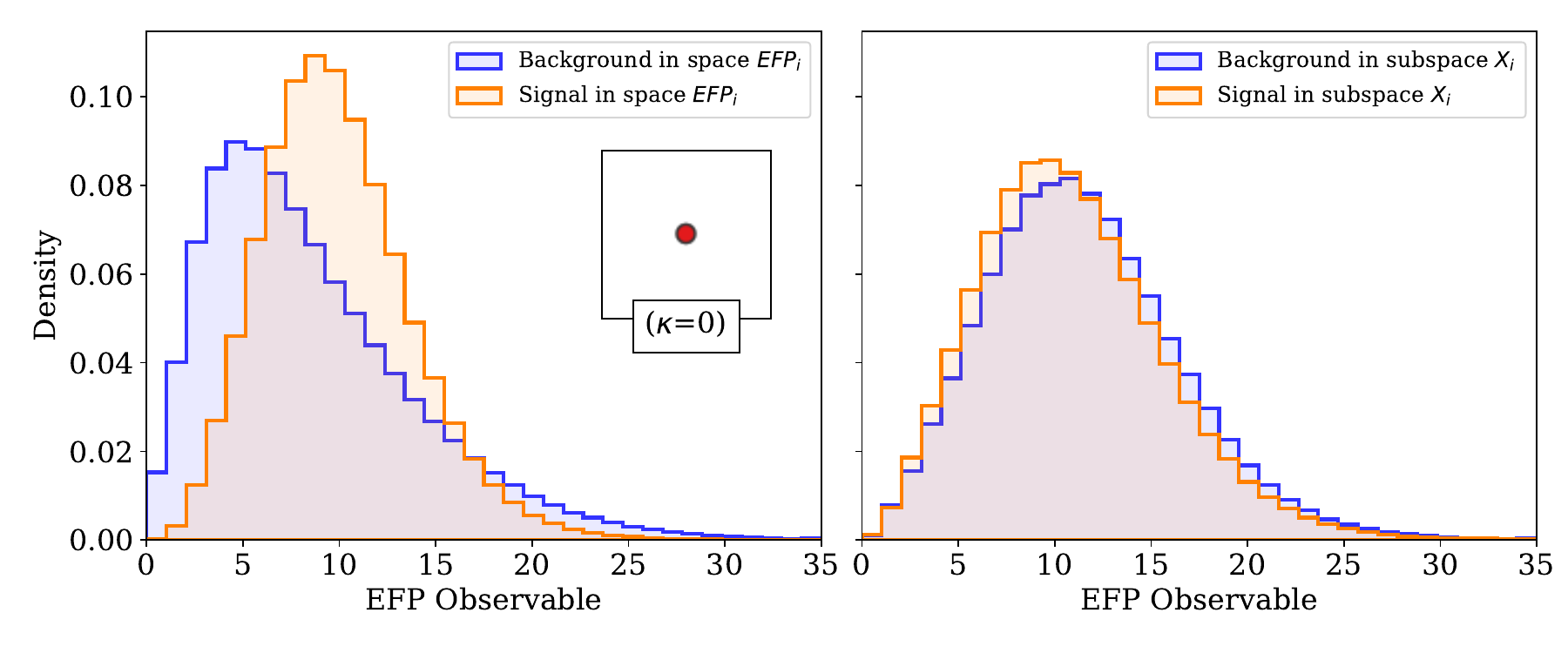}
    \caption{The top three EFPs from the black-box guided search for a seventh HL observable; see Table~\ref{tab:EFPvsBB}. Shown are the EFP distributions for signal and background events, both in the full set of events $X_{\rm all}$ (left column) and in $X_6$ (right column), i.e.\ the differently ordered space between the 6HL and the CNN. The top two observables, while not identical, have very similar functional forms, up to an overall rescaling. The third observable is the jet constituent multiplicity.}
\label{fig:efp_dist}
\end{figure}

For completeness, we show distributions for the top three EFPs from \Tab{tab:EFPvsBB} in \Fig{fig:efp_dist}, both in the full space as well as in $X_6$. The first two observables show good separation between signal and background in the full space, as expected given that their AUC is around $0.8$. The third observable, constituent multiplicity, is a relatively poor discriminant by itself. When restricted to the $X_6$ subspace, there is only modest residual separation power shown by these three observables, but enough to close the performance gap with the CNN.

\subsection{Physics Interpretation}
\label{subsec:supplementing_interpretation}
Our first physics conclusion is that the $\kappa$-augmented EFP space is sufficiently comprehensive to close the performance gap between the 6HL and the CNN. Had we restricted our attention to just the IRC-safe EFPs, this would not have been the case, since the top ranked $\kappa = 1$ EFP in both strategies can only achieve $\AUC = 0.9507$ when combined with the six previous HL observables. Thus, IRC-unsafe information seems essential for closing the performance gap.\footnote{As discussed in \rref{Choi:2018dag}, CNNs are formally IRC safe.  To map this IRC-safe behavior to the EFPs, however, would require very high-point correlators, with in principle as many nodes as pixels in the original jet image.   With IRC-unsafe EFPs, we can instead match the CNN decision boundaries with low-point correlators.}

Fascinatingly, $\kappa = 2$ appears prominently in the top ten EFPs, though in a different form than previously considered in the literature. The key feature of the $\kappa = 2$ EFPs is that they weight higher energy particles more than lower energy particles. Looking at the top $\kappa = 2$ observables in \Tab{tab:EFPvsBB}, they all have the common feature of corresponding to chromatic number $c = 3$ graphs. Chromatic number is the minimum number of colors needed to decorate the nodes of a graph such that no edge connects same-color nodes. If an EFP has chromatic number $c$, then it is only non-zero if the jet has at least $c$ distinct particle directions, making it an effective probe for deviations from $(c-1)$-prong substructure. The $\kappa = 2$ and $c = 3$ EFPs found by our guided strategy therefore probe IRC-unsafe deviations from 2-prong substructure (as one might expect for boosted $W$ tagging), with a particular emphasis on the higher energy particles inside the jet.

By contrast, the only $\kappa = 2$ observable that has received any significant attention in the jet substructure literature is $p_\textrm{T}^D$~\cite{Pandolfi:2012ima,Chatrchyan:2012sn}. In the EFP language, $p_\textrm{T}^D$ is a $c = 1$ graph with no edges:
\begin{equation}
	\ndk{0.045}{1}{0}{0}\ \!\!\!\!\!{}^{(\kappa = 2)} = \sum_{a = 1}^N z^{2}_a \label{eq:pTD}.
\end{equation}
Here, we see that $p_\textrm{T}^D$ is only ranked 5610th by ADO. Apparently, generic IRC-unsafe information is not, by itself, useful for boosted $W$ boson classification, but must be paired with the correct angular dependence to highlight the physics of interest. It is interesting that $\beta = \frac{1}{2}$ is preferred as the angular exponent, since this choice appeared previously in the context of the Les Houches angularity for quark/gluon discrimination~\cite{Gras:2017jty}.

There are also $\kappa = 0$ observables in the top ten EFPs, including the well known constituent multiplicity:
\begin{equation}
	\ndk{0.045}{1}{0}{0}\ \!\!\!\!\!{}^{(\kappa = 0)} = \sum_{a = 1}^N 1 \label{eq:multiplicity}.
\end{equation}
The fact that a $\kappa = 0$, $c = 1$ observable yields nearly the same performance as a class of $\kappa = 2$, $c = 3$ observables is a surprising result of our study. Our tentative interpretation is that this represents two complementary approaches to solving this jet classification task. On the one hand, boosted $W$ bosons are 2-prong objects, so one expects $c = 3$ observables to be most relevant. Indeed, the numerators of \Eqs{eq:C2EFP}{eq:D2EFP} are $c=3$ graphs that probe 2-prong substructure, which was part of the original motivation for the $C_2$ and $D_2$ observables. On the other hand, the background quark and gluon jets are 1-prong objects, and constituent multiplicity is well-known to be a powerful quark/gluon discriminant~\cite{Gras:2017jty} (though sensitive to detector effects~\cite{Kasieczka:2018lwf}). The next $\kappa = 0$ observable on the list has $c = 2$ and $\beta = \frac{1}{2}$, which is an IRC-unsafe probe of 1-prong substructure with an emphasis on collinear physics, which should also be an effective quark/gluon discriminant. This suggests that one can improve the classification performance either with a refined probe of the $W$ boson signal or a refined probe of the quark/gluon background, which happens to have a similar effect on the decision boundaries.

In summary, by translating an ML strategy into a human-readable space, we have identified an important class of jet substructure observables that have been missing from previous boosted $W$ boson studies. This motivates further studies of IRC-unsafe observables, especially high degree EFPs with $\kappa = 2$. In \Sec{sec:discussion}, we discuss the implications of this result on future work with jet substructure observables.

\section{Iteratively Mapping from Minimal Features}
\label{sec:iteratively}
In the previous section, our aim was to supplement an existing set of HL features with one new feature to bridge the gap with the CNN performance. This jet substructure case study is unusual, however, in that it benefits from a highly mature literature of theoretically motivated features. Other applications of our black-box guided strategy may have to begin from a more minimal starting point and build an HL classification strategy essentially from scratch.

In this section, we start from just the most basic jet properties---transverse momentum $p_{\textrm{T}}$ and jet mass $M_\textrm{jet}$---and iteratively identify a small set of EFPs relevant for boosted $W$ boson classification. Using the black-box guided strategy, we are able to match the CNN performance using around 7 EFPs. This is a similar dimensionality to the 7HL combination (which did not include $p_{\textrm{T}}$), though the physics features being probed will turn out to be interestingly different. We then show that this black-box strategy is more computationally efficient than a brute-force search and more effective than a label-guided search.

\subsection{Black-Box Guiding}
\label{subsec:iteratively_blackbox}
Here, we apply the same black-box approach as \Sec{subsec:supplementing_blackbox},  starting just from the smaller set of observables, $p_{\textrm{T}}$ and $M_{\textrm{jet}}$. The motivation for this starting point is as follows. The $W$ boson mass at 80.4 GeV is one of the most important (and obvious) features of boosted $W$ bosons. Because of the choice of $z_a$ variable in \Eq{eq:EFP_z}, though, the EFPs are dimensionless. Therefore, we need at least one dimensionful HL observable to capture the $W$ boson mass peak, and either $p_{\textrm{T}}$ or $M_{\textrm{jet}}$ would suffice for this purpose.

We begin from both $p_{\textrm{T}}$ and $M_{\textrm{jet}}$ for two reasons. The first is that they are ubiquitous jet observables appearing in myriad collider studies. The second is to streamline the selection of EFPs. Naively, $M_{\textrm{jet}}$ could be derived from $p_{\textrm{T}}$ using the EFP in \Eq{eq:EFP_theta} with $\kappa =1$ and $\beta = 2$:
\begin{equation}
\label{eq:mass_approx}
\ndk{0.045}{2}{1}{0} {}^{(\kappa = 1, \beta = 2)} \approx \frac{M^2_{\textrm{jet}}}{{p^2_{\textrm{T}}}}.
\end{equation}
With the choice of $\theta_a$ variable in \Eq{eq:EFP_theta}, though, \Eq{eq:mass_approx} is only approximately true, so multiple EFPs are needed to map out the mass information if $p_{\textrm{T}}$ is the only dimensionful scale. We checked that the black-box strategy is still effective starting from just $p_{\textrm{T}}$ or from just $M_{\textrm{jet}}$, but the chosen EFPs tend to be more mass-like in their structure. By contrast, starting from both $p_{\textrm{T}}$ and $M_{\textrm{jet}}$ yields more variation in the types of EFPs selected.

\begin{figure}[t]
	\centering
	\includegraphics[width=0.48\textwidth]{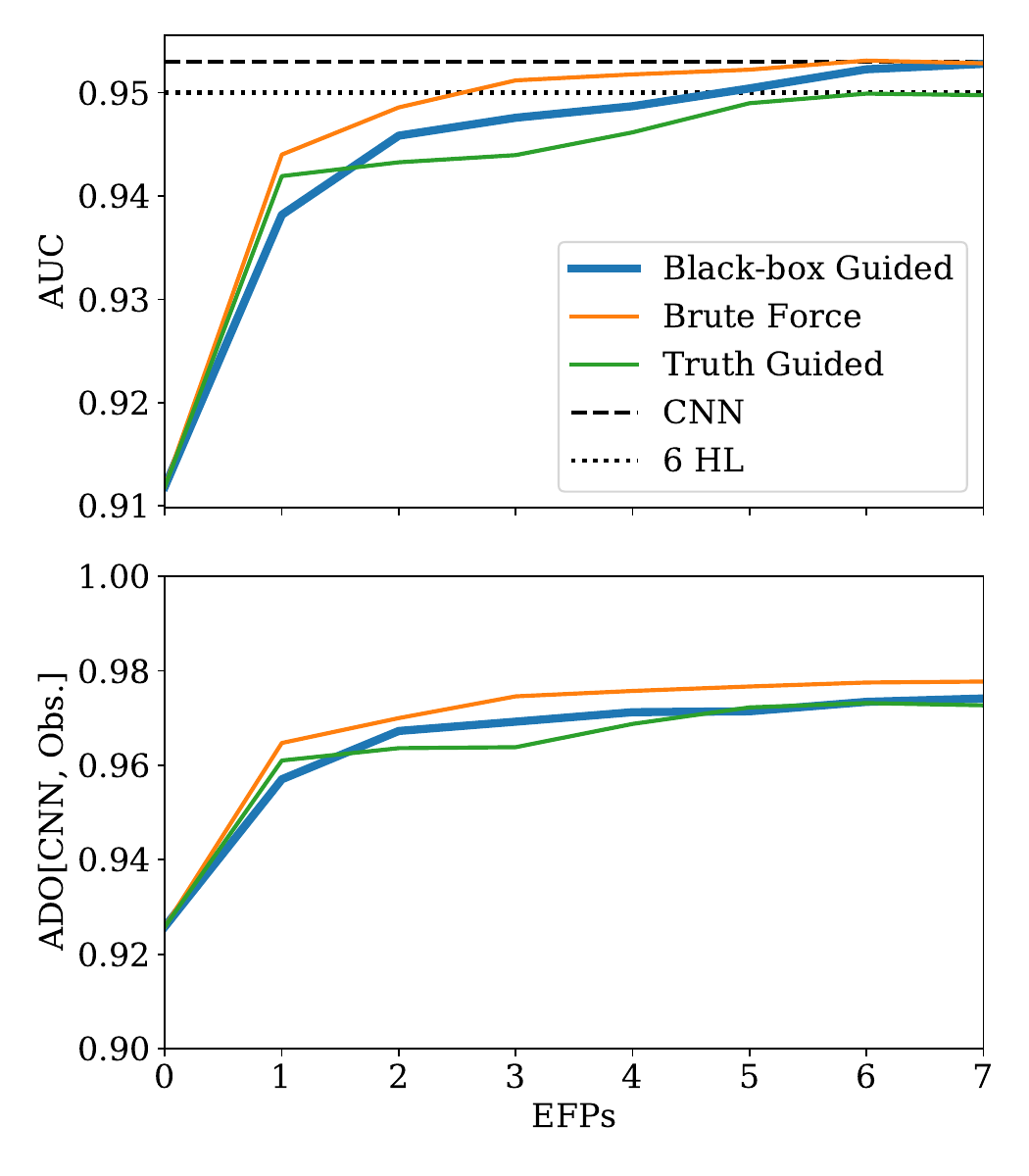}
	\caption{
	Performance of the black-box guided search strategy to map the CNN solution into human-interpretable observables. Here, we start from just the basic jet features $p_{\textrm{T}}$ and $M_{\textrm{jet}}$ and iteratively add one EFP at a time. The performance is shown in terms of AUC (top) and ADO (bottom) as a function of the scan number. The performance of a brute-force scan of the EFP space (\Sec{subsec:iteratively_bruteforce}) and a truth-label guided search (\Sec{subsec:iteratively_truthlabels}) are also shown. For reference, the performance of the CNN and of the existing 6HL features are indicated by horizontal lines.}
	\label{fig:maxGraph}
\end{figure}

\begin{figure}[t]
	\centering
	\includegraphics[width=0.48\textwidth]{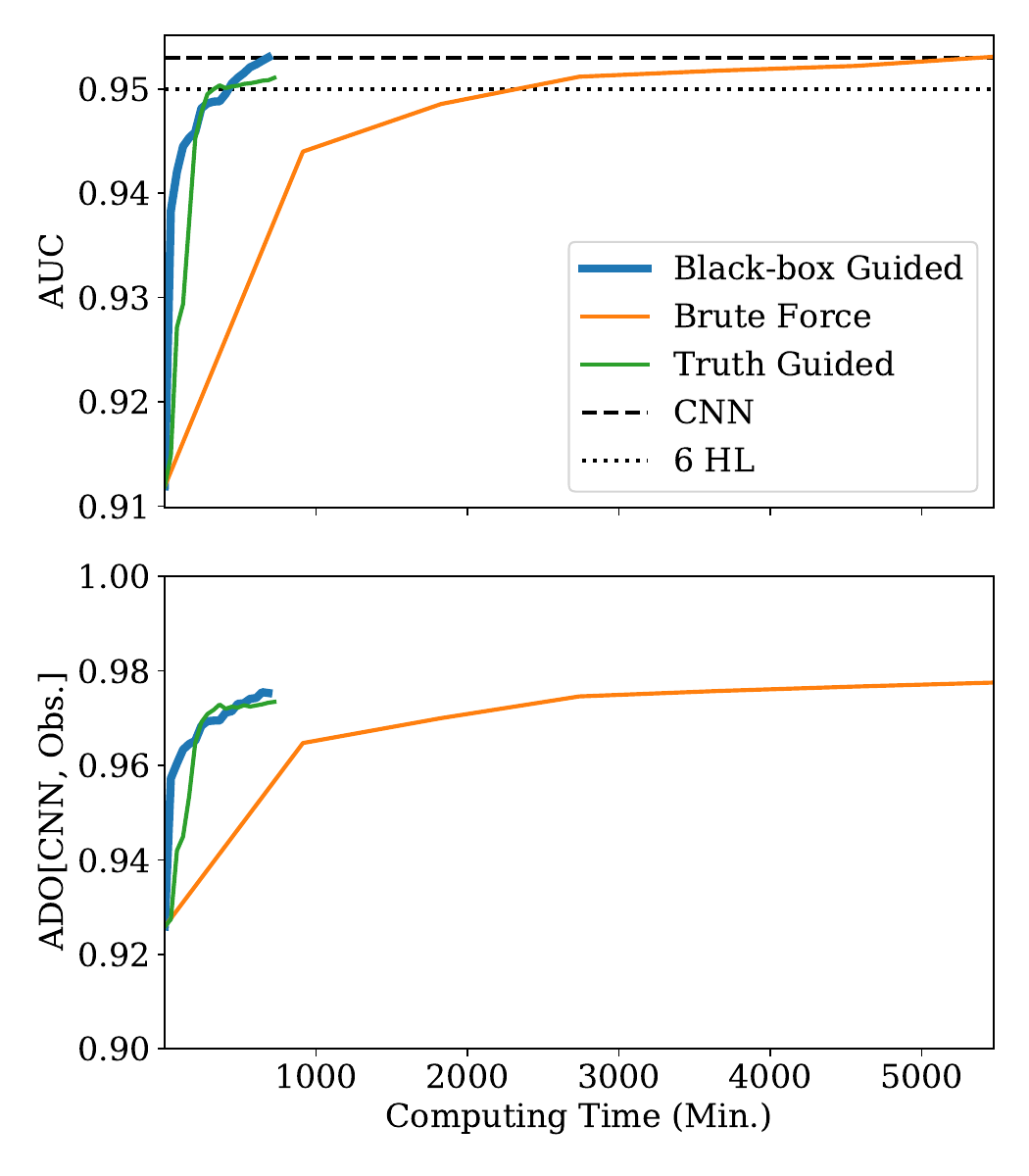}
	\caption{The same as \Fig{fig:maxGraph}, but now plotted in terms of the cumulative computing time. 
	}
	\label{fig:maxComputeGraph}
\end{figure}

We start by training an NN on just the $p_{\textrm{T}}$ and $M_{\textrm{jet}}$ information:   
\begin{equation}
\label{eq:PTM}
\PTM \equiv \NN \big[p_{\textrm{T}},M_{\textrm{jet}} \big].
\end{equation}
This yields an AUC of $0.9119$, which is substantially below the CNN performance for boosted $W$ boson tagging. We then restrict our attention to the subset of events that are differently ordered by these minimal features relative to the CNN:
\begin{equation}
	X_0 = \Big\{ (x,x') \, \Big| \, \DO\big[\CNN, \PTM \big](x,x') = 0 \Big\}.
\end{equation}
The ADO between the CNN and $\PTM$ is 0.9150, so $X_0$ contains 8.5\% of the original $X_{\rm all}$ sample, though we only consider a random subset of $5 \times 10^7$  pairs in $X_0$ to reduce the computational burden. Our aim is to find a set of EFPs that can order these signal and background events the same as the CNN decision boundaries.

To identify the $n$-th EFP, we use the black-box guided strategy of \Sec{subsec:strategy_blackbox}, adapted to the current notation:
\begin{equation}
	\EFP_n = \argmax_{\EFP \in \mathcal{S}} \ADO[\CNN, \EFP]_{X_{n-1}}.
\end{equation}
We construct a new joint classifier that includes this EFP: 
\begin{equation}
\label{eq:nEFP_network}
	\text{HLN}_n  \equiv \NN[p_{\textrm{T}},M_{\textrm{jet}}, \EFP_1, \ldots, \EFP_n].
\end{equation}
This allows us to identify the remaining differently-ordered subset of events:
\begin{equation}
	X_n = \Big\{ (x,x') \, \Big| \, \DO[\CNN,\text{HLN}_n](x,x') = 0 \Big\},
\end{equation}
where in each iteration we only keep a random subset of $5 \times 10^7$ pairs. The main computational cost in this procedure is in training the joint classifier in \Eq{eq:nEFP_network}.

\begin{table*}
\centering
\begin{tabular}{c|c|ccc|cc|cc}
	\hline \hline
	Iteration ($n$) & EFP & $\kappa$ & $\beta$ & Chrom \# & $\text{ADO}[\text{EFP}, \text{CNN}]_{X_{n-1}}$ & AUC[EFP] & $\text{ADO}[\text{HLN}_n,\text{CNN}]_{X_{\rm all}}$ & $\text{AUC}[\textrm{HLN}_n]$\\ \hline \hline
	&&&&&&&& \\
0 & $M_{\rm jet} + p_{\textrm{T}}$ & -- &-- & -- & --& --& 0.9259 & 0.9119\\[0.6em]
1 & $\tndk{5}{5}{4}$ & 2 & $\frac{1}{2}$ & 2 & 0.8144 & 0.8190 & 0.9570 & 0.9382 \\
2 & $\tndk{2}{4}{0}$ & 0 & 2 & 2 & 0.6377 & 0.8106 & 0.9673 & 0.9458 \\
3 & $\tndk{1}{0}{0}$ & 0 & -- & 1 & 0.5460 & 0.6737 & 0.9692 & 0.9476 \\
4 & $\tndk{8}{7}{0}$ & 1 & $\frac{1}{2}$ & 2 & 0.5274 & 0.8464 & 0.9712 & 0.9487 \\
5 & $\tndk{1}{0}{0}$ & $-1$ & -- & 1 & 0.5450 & 0.5882 & 0.9714 & 0.9504 \\
6 & $\tndk{4}{6}{21}$ & 1 & $\frac{1}{2}$ & 4 & 0.5382 & 0.7678 & 0.9734 & 0.9523 \\
7 & $\tndk{6}{5}{0}$ & $-1$ & $\frac{1}{2}$ & 2 & 0.5561 & 0.5957 & 0.9741 & 0.9528 \\
	\hline \hline
\end{tabular}
\caption{The EFPs selected during each iteration of the black-box guiding strategy beginning from HLN$_0$, which uses just $p_\textrm{T}$ and $M_\textrm{jet}$.  For each iteration, the selected EFP is the one with the largest ADO with the CNN in the differently-ordered subspace $X_{n-1}$.}
\label{tab:iterative}
\end{table*}

The AUC and ADO values for this black-box guided procedure are shown in \Fig{fig:maxGraph} versus EFP scan iteration, and in \Fig{fig:maxComputeGraph} versus cumulative computing time. More details about the selected EFPs are given in \Tab{tab:iterative}. By the 5th iteration, the AUC performance matches that of the 6HL combination. By the 7th iteration, the AUC performance matches that of the CNN, with an ADO of 0.974, indicating closer agreement with the CNN decisions than we found with the $7\HL_{\text{black-box}}$  strategy. Since we started from a minimal set of jet features, it is not surprising that the EFPs identified here are qualitatively different from the ones in \Sec{sec:supplementing}. The physics interpretation of these various EFPs will be presented in \Sec{subsec:iteratively_physics}.

\subsection{Comparison to Brute Force Search}
\label{subsec:iteratively_bruteforce}
An alternative approach to maximizing ADO is to perform a brute-force search through the space of EFPs to find the set that maximally matches the decisions of the CNN. This is much more computationally expensive than the black-box guided strategy, but it has the potential to converge to a smaller number of EFPs if there are important correlations between the observables. In an absolute brute force search, one would construct all possible sets of EFPs, and evaluate the ADO of each relative to the CNN; given the number of graphs and combinations, this would be completely intractable. Instead, we attempt an iterative greedy algorithm, which incrementally builds the EFP set. This is still computationally expensive, but (borderline) tractable.

We again start from the jet $p_{\textrm{T}}$ and $M_{\textrm{jet}}$ information, but immediately train a joint classifier using each of the EFPs as an input:
\begin{equation}
\NN\Big[p_{\textrm{T}}, M_{\textrm{jet}}, \EFP \Big].
\end{equation}
We then select the EFP that yields the largest ADO with the CNN, evaluated on the full training set.In the first iteration, we select the EFP via:
\begin{equation}
\EFP_1 = \argmax_{\EFP \in \mathcal{S}} \ADO\Big[\CNN, \NN\big[p_{\textrm{T}}, M_{\textrm{jet}}, \EFP \big]\Big]_{X_{\rm all}}.
\end{equation}
We repeat this procedure in each subsequent iteration, choosing the EFP that yields the largest improvement in ADO when combined with the previous observables:
\begin{equation}
\label{eq:bruteforcescan}
\EFP_n = \argmax_{\EFP \in \mathcal{S}} \ADO\Big[\CNN, \NN\big[\ldots, \EFP_{n-1}, \EFP \big]\Big]_{X_{\rm all}}.
\end{equation}
The key difference from the black-box guided strategy is that the joint classifier is trained before evaluating the ADO, and the ADO is evaluated on the full training set, instead of just the differently-ordered subset.

The primary computational cost of the brute force approach comes from training the joint classifier appearing in \Eq{eq:bruteforcescan}, which combines each EFP with the current set of observables. This has to be done for each EFP under consideration, and it is too computationally expensive to examine all 7,545 EFP graphs over multiple iterations. Therefore, we only consider a subset of graphs at each iteration, which means there is no guarantee that the brute force method will perform better than the black-box guided strategy. For our purposes, our subset consists of the 54 connected graphs of degree $d\leq 5$ and $(\kappa, \beta)$ choices of $\kappa=[\frac{1}{2}, 1, 2]$ and $\beta=[\frac{1}{2}, 1, 2]$. This reduces our original search space down to a more manageable 486 choices.

The results from this brute force procedure are shown in  \Fig{fig:maxGraph} in terms of the ADO and AUC values after each iteration. In the first few iterations, the AUC and ADO values are higher than for black-box guiding, achieving a comparable performance to the original 6HL result after the inclusion of a third EFP. The brute force process continues until it matches the CNN performance with 6 EFPs (8 HL inputs total). As one would expect, the brute force approach performs well as it is effectively trying every possible combination of inputs and selecting the best. This computational cost, however, must be weighed against the marginal decrease in the number of EFPs required to match the CNN as well as the need to restrict the input space prior to exploring the performance. As shown in \Fig{fig:maxComputeGraph}, the brute force approach does not complete even a single iteration before the guided approaches have converged to a complete solution. 

Finally, for completeness, we consider the alternative brute-force approach of a network trained with all available EFPs. Using the baseline DNN architecture in \App{app:dnn_ll}, a network was trained with $M_{\textrm{jet}}$, $p_\textrm{T}$, and the reduced set of 486 EFPs as input features.  The performance of ``488HL'' is shown in \Tab{tab:AUC6HL}, with marginally better performance than the CNN, indicating that the EFPs are effectively a complete basis for this task.

\subsection{Comparison to Truth-Label Guiding}
\label{subsec:iteratively_truthlabels}
In the black-box guided strategy, the CNN and the ADO similarity metrics are auxiliary tools to help identify a set of EFPs that maximizes the classification performance. Assuming the EFP space is sufficiently complete and labeled samples exist, one could dispense with the CNN entirely and simply search the space of EFPs for the most powerful set that maximizes AUC, in an approach similar to that of \rref{Datta:2017lxt}.  As a computationally tractable alternative to the brute force search, we test what happens when the selection of the EFP is guided by the truth labels, instead of by the ADO.

Analogously to decision ordering in \Eq{eq:DO_def}, we define \emph{truth ordering} (TO) for a pair of signal/background points $x$ and $x'$ and a decision function $f$:
\begin{equation}
	\TO[f](x,x') = \Theta \big( f(x) - f(x') \big),
\end{equation}
where 1 corresponds to $f$ correctly ordering the points and 0 corresponds to inverted ordering. Starting again from the jet $p_{\textrm{T}}$ and $M_{\textrm{jet}}$ information, we identify the subset of event pairs that are incorrectly ordered: 
\begin{equation}
	Y_0 = \Big\{ (x,x') \, \Big| \, \TO\big[\PTM \big](x,x') = 0 \Big\}.
\end{equation}
In each iteration, we find the EFP that has the highest AUC in the incorrectly-ordered subspace,
\begin{equation}
	\EFP_n = \argmax_{\EFP \in \mathcal{S}} \AUC[\EFP]_{Y_{n-1}},
\end{equation}
construct a new joint classifier HLN$_n \equiv \PTM + n\EFP$, and identify the next incorrectly-ordered subset of events:
\begin{equation}
	Y_n = \Big\{ (x,x') \, \Big| \, \TO[\text{HLN}_n](x,x') = 0 \Big\}.
\end{equation}
Note that this procedure is completely independent of the CNN.

The results from this truth-label guided procedure are shown in in \Fig{fig:maxGraph} in terms of the AUC and ADO. In the first iteration, the classification performance is better than in the black-box guided search, which makes sense since the label guided method is trying to optimize AUC directly. After 7 iterations, though, the classification performance never rises above $\AUC = 0.951$. As mentioned in \Sec{subsec:strategy_blackbox}, isolating the incorrectly-ordered pairs turns out to be counter productive, since some of these pairs could never be ordered correctly even by the optimal classifier. This emphasizes the value of using the ADO relative to an already-trained network, to make sure attention is focused on event pairs that have a chance to be correctly ordered.

\subsection{Physics Interpretation}
\label{subsec:iteratively_physics}
\begin{figure}[t]
	\centering
	\includegraphics[width=0.48\textwidth]{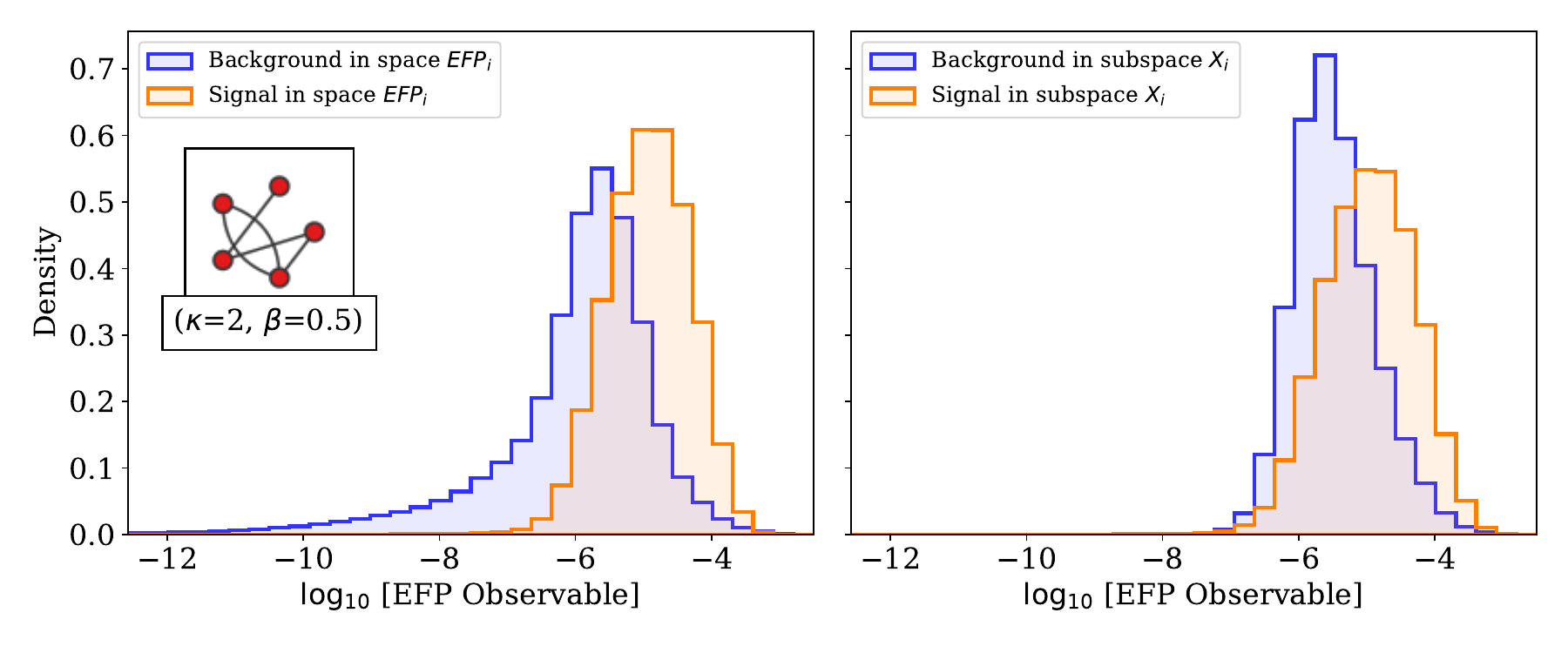}
	\includegraphics[width=0.48\textwidth]{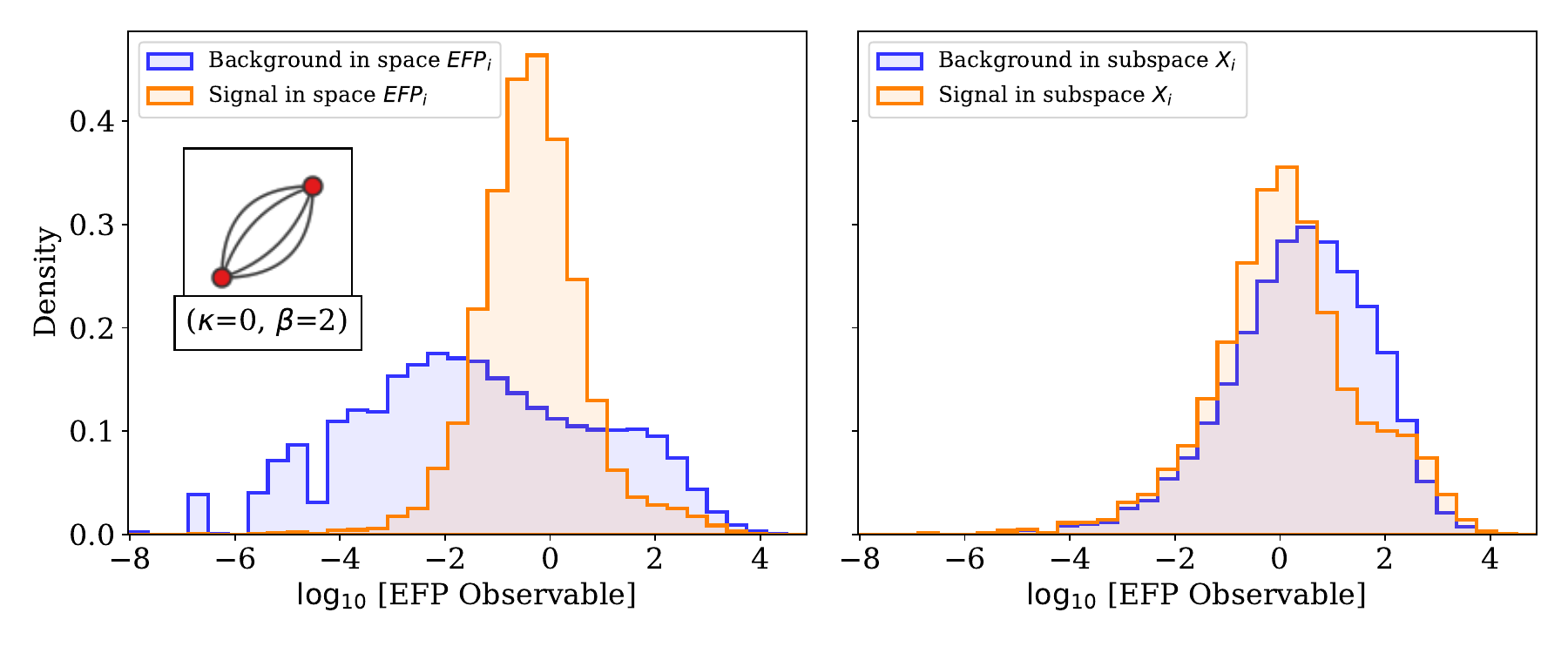}
	\includegraphics[width=0.48\textwidth]{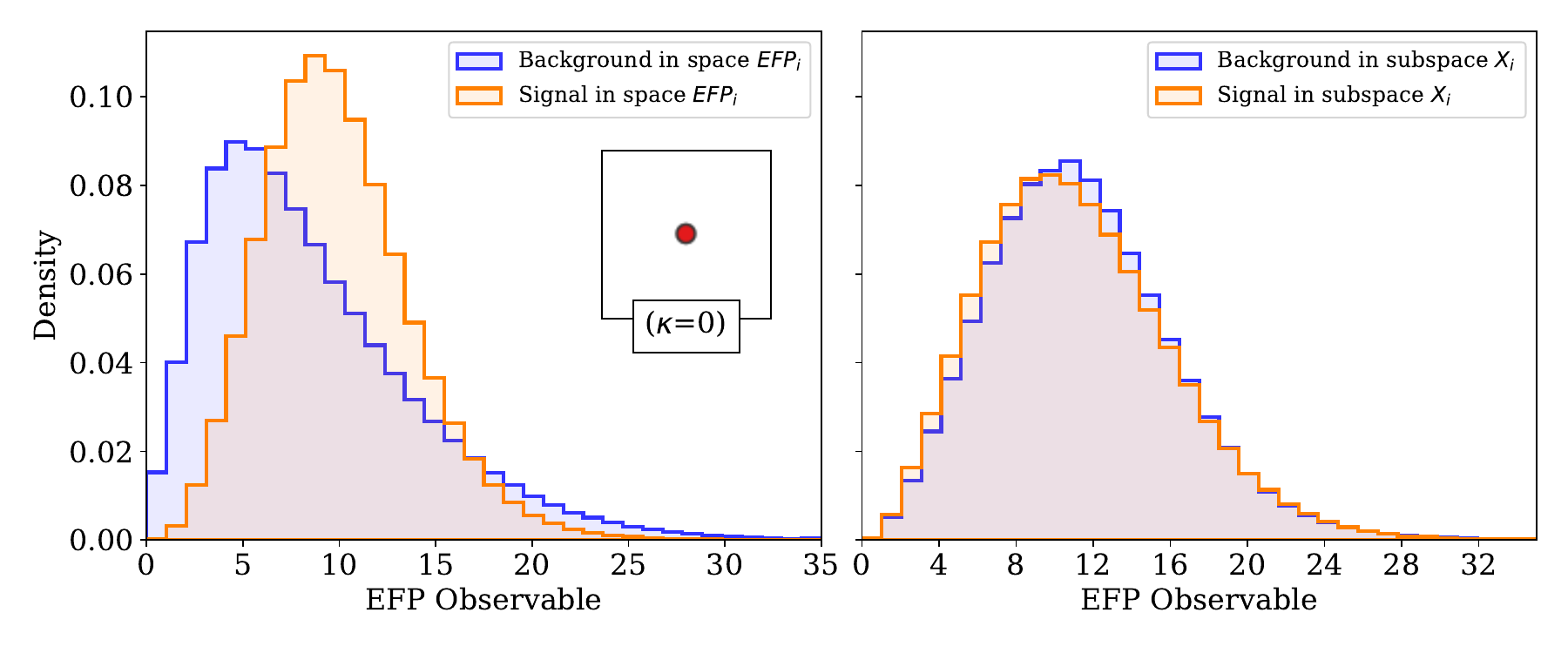}
	\includegraphics[width=0.48\textwidth]{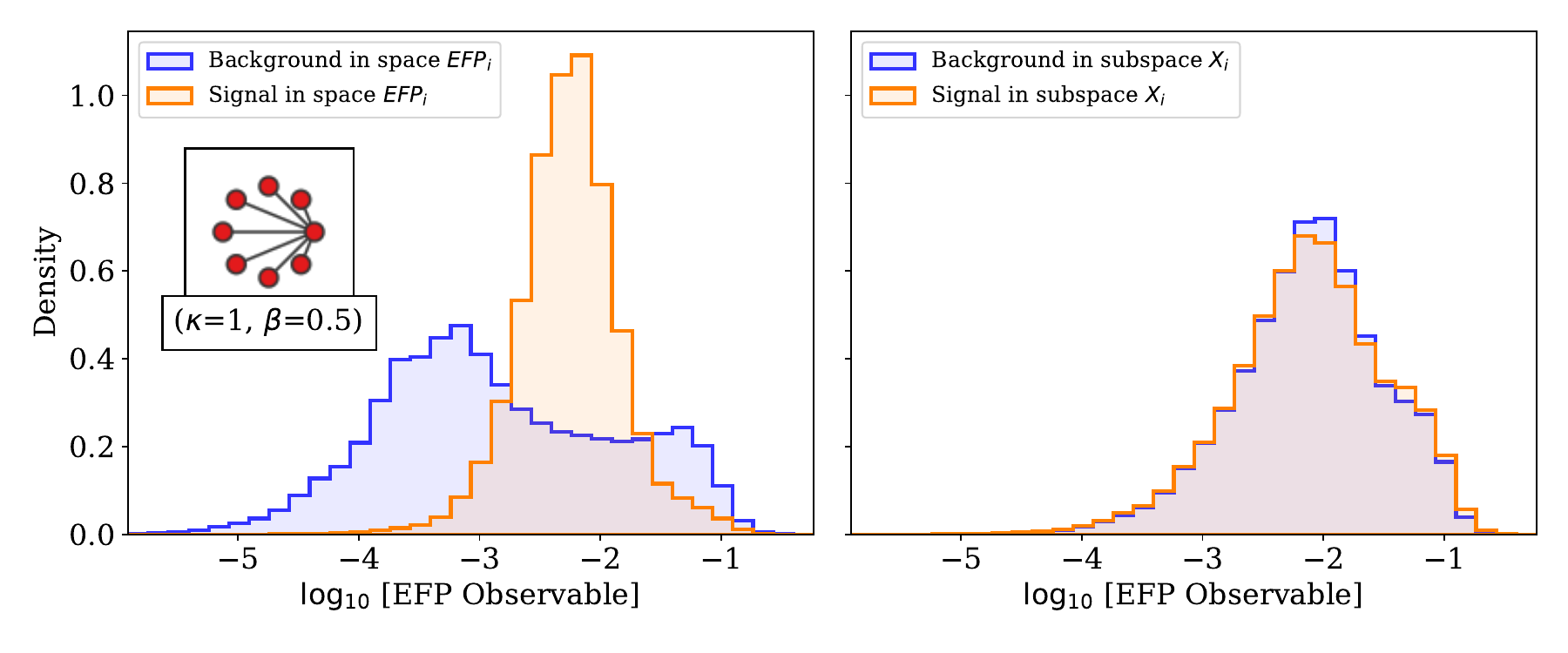}
    \caption{The first four EFP graphs selected by the black-box guided strategy beginning from the minimal set of HL observables, $p_{\textrm{T}}$ and $M_{\textrm{jet}}$; see Table~\ref{tab:iterative}. Shown are the EFP distributions for signal and background events, both in the full set of events $X_{\rm all}$ (left column) and in $X_n$ (right column), i.e.\ the differently ordered space between the HLN$_n$ and the CNN after $n$ iterations.}
\label{fig:efp_dist_iterative}
\end{figure}

By translating the CNN into a space of physically-motivated observables, we can gain physical insight into the observables used in the classification decision. In particular, the first few observables in \Tab{tab:iterative} give us a glimpse at a possible alternative history for the field of jet substructure, if combinations like $C_2$ and $D_2$ had not been previously identified. Distributions of the EFPs found in the first four iterations are shown in \Fig{fig:efp_dist_iterative}.

After $p_{\textrm{T}}$ and $M_{\textrm{jet}}$, the first EFP selected by the black-box guided strategy is:
\begin{equation}
	\label{eq:g116}
	\ndk{0.045}{5}{5}{4} {}^{\left(\kappa = 2, \beta = \tfrac{1}{2} \right)}.
\end{equation}
The fact that a $\kappa = 2$ observable shows up early in the iterative procedure bolsters the evidence from \Sec{subsec:supplementing_blackbox} that these kinds of observables are important for mapping the CNN strategy. This is a chromatic number $c = 2$ graph, so just like jet mass, it probes deviations from 1-prong substructure. However, it uses a 5-point correlator (unlike mass which is a 2-point correlator) and it uses the $\beta = \frac{1}{2}$ angular exponent (unlike mass which uses $\beta = 2$). Putting these together, \Eq{eq:g116} is an IRC-unsafe probe of hard, small-angle radiation.

The second EFP is also IRC unsafe and also corresponds to a $c = 2$ graph:
\begin{equation}
	\label{eq:g84}
	\ndk{0.045}{2}{4}{0} {}^{(\kappa = 0, \beta = 2)}.
\end{equation}
Here, though, we have $\kappa = 0$ and $\beta = 2$, which is a probe of soft, wide-angle radiation. It is interesting that the black-box guided strategy selects these two complementary $c = 2$ observables in the first two iterations, indicating the importance of 1-prong substructure probes even if the goal is to identify 2-prong boosted $W$ bosons. 

The third EFP is constituent multiplicity, as seen before in \Eq{eq:multiplicity}, which reinforces the idea that controlling the composition of the quark/gluon background is important for $W$ tagging. These three observables, together with $p_{\textrm{T}}$ and $M_{\textrm{jet}}$, yield an AUC of 0.9476. This is not as good as the 6HL combination, but still quite encouraging given that we did not give the black-box guided strategy any information about the ratio structures used to construct $C_2$ and $D_2$.

The main surprise from this study is that IRC-safe information was not selected by the black-box guided search until the fourth iteration:
\begin{equation}
	\label{eq:fourth}
	\ndk{0.045}{8}{7}{0} {}^{\left(\kappa = 1, \beta = \frac{1}{2}\right)}.
\end{equation}
Moreover, it is a $c= 2$ graph, so still a probe of 1-prong substructure. Only in interaction six do we finally see a higher chromatic number graph, but the guided search skips over the $C_2/D_2$-like graphs with $c = 3$ and goes straight to $c = 4$. The black-box guided strategy has identified a very different strategy for boosted $W$ boson tagging that nevertheless matches the 6HL combination with a comparable number of observables.

One interpretation of this result is that it simply reflects the ``entropy'' of our HL space. There are 4 times as many IRC-unsafe observables in our HL collection than IRC-safe ones, so just by random chance, one expects to see more unsafe observables in the scan. Indeed, there are IRC-safe observables that are highly ranked in the first three iterations, just not at the top of the list. Another interpretation is that the black-box guided strategy is teaching us that IRC-unsafe information is more relevant for boosted $W$ tagging than one might naively think. A related observation was made in \rref{Buckley:2020kdp}, which introduced a color ring observable to identify color-singlet configurations. Intriguingly, when restricted to three particles, the angular structure of \Eq{eq:g116} defines similar decision boundaries to the color ring.\footnote{We thank Andrew Larkoski for discussions of this point.} Either way, by searching through a large space of HL observables in a systematic way, the black-box guided strategy has given us a new perspective on an old problem in a human-readable format.

\section{Discussion}
\label{sec:discussion}
The ever increasing complexity of new ML strategies has produced better classification performance for various physics problems. At the same time, the increasing opaqueness of these methods has widened the gap between our understanding of a problem and our appreciation of the ML solution. In this paper, we have proposed a new technique for mapping an ML solution into a space of human-interpretable observables. Our guided strategies mitigate some of the well-founded concerns about black-box approaches, while still allowing us to capitalize on the black-box performance to efficiently guide the selections of HL observables. The end result is a set of HL observables that have a more direct physical interpretation and allow for a more transparent treatment of systematic uncertainties.

In our jet substructure case study, we have shown that the black-box guided strategy could be used to isolate information that is not captured by previous HL representations. Remarkably, only a single observable was needed to close the performance gap identified in \rref{Baldi:2016fql}, nearly duplicating the CNN strategy with a low-dimensional input representation. Beginning from a minimal set of basic jet observables ($p_\textrm{T}$ and $M_{\textrm{jet}}$), we successfully condensed the CNN behavior to a small set of EFP observables which reproduce its performance and very nearly match its decisions.  It would be interesting to study the utility of the EFPs in more complicated contexts, such as event-wide classification tasks.

Interestingly, the structure of the selected EFPs differ in qualitative ways from the $C_2$ and $D_2$ jet substructure observables custom designed for boosted $W$ boson classification. While these previous observables are based on fully connected graphs, the guided strategy picked out multi-node EFP graphs with relatively low chromatic number. While these previous observables use the IRC-safe choice of $\kappa = 1$, the guided strategy emphasized the importance of unsafe $\kappa = 2$ observables, particularly ones with non-trivial angular dependence. This motivates further physics studies of these exotic EFPs. It is worth emphasizing that we were only able to identify these new observables because we considered a sufficiently large space of HL observables. There may be other hidden organizing principles to exploit for jet substructure studies, which motivates the construction of alternative sets of observables based on different physical principles than the EFPs. In particular, we did not capitalize on the power counting and scaling properties of ratio/product observables \cite{Larkoski:2014gra,Moult:2016cvt,Datta:2017lxt,Datta:2019ndh,Sirunyan:2020lcu}, which may reveal more efficient HL observables for jet classification. It may also be beneficial to leverage first-principles knowledge about signal/background likelihood ratios~\cite{Soper:2011cr,Soper:2012pb,Soper:2014rya,FerreiradeLima:2016gcz,Larkoski:2019nwj,Buckley:2020kdp,Kasieczka:2020nyd} to identify promising HL observables. 

These results have important implications for what we should regard as ``best practices'' in the application of ML methods to high-energy physics problems. Primarily, we should be more wary of utilizing opaque ML strategies which obscure how a problem is solved in exchange for relatively small classification improvements. In general, one should evaluate whether ``additional'' information captured by DNNs represents genuine patterns or is the byproduct of something unintentionally pressed into the data during simulation and then re-discovered by the network.

The informational gap in our benchmark problem could be closed using a single HL observable, suggesting that the CNN strategy was not relying on subtle correlations among the low-level  features, but rather exploiting information encodable into a $\kappa = 2$ EFP. Thus, instead of a purely performance-oriented approach, we suggest a strategy of using deep networks to establish performance benchmarks, but always seek to translate ML strategies into a more tractable space of well-motivated physical observables. If this proves to be impossible or impractical, it might be that the ML approach really is identifying genuinely new information, or more likely, that the space of physical observables needs to be augmented or optimized. 

More broadly, our view is that the ultimate goal of ML research in high-energy physics should not be to develop artificial-intelligence physicists which (or should we say \emph{who}?) can blindly process raw data and make statements about the Universe without being able to communicate the intermediate steps. The power of modern ML can certainly be used to identify gaps in our knowledge where existing human-engineered approaches are insufficient. At the end of the day, though, we should insist that data analysis strategies used to make statements about physics should be understandable to human physicists.

\begin{acknowledgments}
	We thank Pierre Baldi, Timothy Cohen, Michael Fenton, Patrick Komiske, Andrew Larkoski, Eric Metodiev, Benjamin Nachman, Tilman Plehn, Peter Sadowski, and Edison Weik for helpful conversations and feedback. JT was supported by the U.S.\ Department of Energy (DOE), Office of High Energy Physics under Grant No. DE-SC0012567. DW and TF are supported by the U.S.\ Department of Energy (DOE), Office of Science under Grant No. DE-SC0009920. TF was supported by the National Science Foundation under Award No. 1633631.
\end{acknowledgments}

%%% Appendix %%%
\appendix

\section{Network Architectures and Hyperparameters}
\label{app:casestudy_nn}

For consistency, all neural networks, regardless of input data type, use a set of common settings and training procedures:
\begin{itemize}
	\item $N = 5\times 10^6$ training samples, broken down as:
	\begin{itemize}
		\item 70\% training,
		\item 15\% validation, 
		\item 15\% testing.
	\end{itemize} 
	\item Training samples are pre-processed via \textsc{sci-kit}'s \texttt{StandardScaler}.
	\item Adam optimizer used with default \textsc{tensorflow} settings: 
	\begin{itemize}
		\item \verb@learning_rate = 0.001@,
		\item \verb@beta1 = 0.9@,
		\item \verb@beta2 = 0.999@,
		\item \verb@epsilon = 1e-07@,
		\item \verb@amsgrad = False@.
	\end{itemize}
	\item Output layer uses a \verb@sigmoid@ activation.
	\item Uncertainties are calculated via a 10-fold cross-validation (see \App{app:10-fold}).
	\item Early stopping on \verb@validation_loss@ with a patience of 30 epochs.
	\item Model checkpoints saved (for best results only) on minimum validation loss.
\end{itemize}

\subsection{Baseline Convolutional Neural Network}
 \label{app:dnn_ll}
  
 The following training details are specific to all convolutional neural networks trained on jet-images:
 \begin{itemize}
 	\item Input data undergoes a log transformation prior to \texttt{StandardScaler} pre-processing.
 	\item Hidden layers consist of 5 hidden convolutional layers with the parameters:
 	\begin{itemize}
 		\item 500 nodes,
 		\item \verb@kernel_size = (4,4)@,
 		\item \verb@strides = (1,1)@,
 		\item \verb@padding = 'valid'@,
 		\item \verb@kernel_initializer = 'glorot_normal'@,
 		\item \verb@activation = 'relu'@,
 		\item \verb@kernel_constraint = max_norm(3)@.
 	\end{itemize}
 	\item 3 dropout layers (i.e.\ 1 between each convolutional layer) with a rate of 0.2.
 \end{itemize}

\subsection{Baseline Dense Neural Network}
 \label{app:dnn_hl}
  
The following training details are specific to all dense neural networks acting on jet substructure observables, including EFPs:
\begin{itemize}
	\item Hidden layers consist of 3 hidden dense layers with the parameters:	
	\begin{itemize}
		\item 300 nodes,
		\item \verb@activation = 'relu'@,
		\item \verb@kernel_constraint = max_norm(3)@.
	\end{itemize}
	\item 2 dropout layers (i.e.\ 1 between each dense layer) with a rate of 0.5.
\end{itemize}

\subsection{K-fold Validation}
\label{app:10-fold}

To derive uncertainties on the trained model prediction accuracy, we use the bootstrap cross-validation package in \textsc{sci-kit} to equally divide the test set 10 times and measure the performance across 10 bootstrapping iterations.
Averages and standard deviations are then taken from these 10 iterations to define the central value and uncertainties of the AUC.

%%% Bibliography %%%
\bibliography{main}

\end{document}